\documentclass[sigconf]{acmart}

\usepackage[linesnumbered,ruled,vlined]{algorithm2e}
\usepackage[svgnames,table]{xcolor} 

\SetKwComment{Comment}{// }{}
\SetCommentSty{mycommentstyle}

\usepackage{graphicx}
\usepackage{tikz}
\usepackage{tikz-cd}
\usepackage[caption=false]{subfig}
\usepackage{pgf}
\usepackage{pgfplots}
\pgfplotsset{compat=1.18}
\usepackage{pgfplotstable}
\usepgfplotslibrary{patchplots}
\usepgfplotslibrary{colormaps}
\usepackage{multirow}
\usepackage{multicol}
\usepackage{makecell}

\usepackage{amsmath}
\DeclareMathOperator\supp{supp}
\newcommand*\D{\mathop{}\!\mathrm{d}}

\usepackage[english]{babel}
\usepackage{amsthm}

\usepackage{booktabs}
\theoremstyle{plain}

\newtheorem{theorem}{Theorem}[section]
\newtheorem{lemma}[theorem]{Lemma}

\newtheorem{proposition}[theorem]{Proposition}

\newtheorem{assumption}[theorem]{Assumption}

\newtheorem{remark}[theorem]{Remark}

\AtBeginDocument{%
  }

\setcopyright{acmlicensed}
\copyrightyear{2018}
\acmYear{2018}
\acmDOI{XXXXXXX.XXXXXXX}
\acmConference[Conference acronym 'XX]{Make sure to enter the correct
  conference title from your rights confirmation email}{June 03--05,
  2018}{Woodstock, NY}
\acmISBN{978-1-4503-XXXX-X/2018/06}

\acmSubmissionID{1182}

\begin{document}

\title{Noise Reduction for Pufferfish Privacy: A Practical Noise Calibration Method}

\author{Wenjin Yang}
\email{wenjinyang@bit.edu.cn}
\affiliation{%
  \institution{Beijing Institute of Technology}
  \city{Beijing}
  \country{China}
}

\author{Ni Ding}
\email{dingni529@gmail.com}
\affiliation{%
  \institution{University of Auckland}
  \city{Auckland}
  \country{New Zealand}
}

\author{Zijian Zhang}
\email{zhangzijian@bit.edu.cn}
\affiliation{%
  \institution{Beijing Institute of Technology}
  \city{Beijing}
  \country{China}
}

\author{Jing Sun}
\email{jing.sun@auckland.ac.nz}
\affiliation{%
  \institution{University of Auckland}
  \city{Auckland}
  \country{New Zealand}
}

\author{Zhen Li}
\email{zhen.li@bit.edu.cn}
\affiliation{%
  \institution{Beijing Institute of Technology}
  \city{Beijing}
  \country{China}
}

\author{Yan Wu}
\email{wuyan.bit@gmail.com}
\affiliation{%
  \institution{Beijing Institute of Technology}
  \city{Beijing}
  \country{China}
}

\author{Jiahang Sun}
\email{sunjh@bit.edu.cn}
\affiliation{%
  \institution{Beijing Institute of Technology}
  \city{Beijing}
  \country{China}
}

\author{Haotian Lin}
\email{hlin228@aucklanduni.ac.nz}
\affiliation{%
  \institution{University of Auckland}
  \city{Auckland}
  \country{New Zealand}
}

\author{Yong Liu}
\email{liuyong03@qianxin.com}
\affiliation{%
  \institution{Qi An Xin Technology Group Inc.; Zhongguancun Laboratory}
  \city{Beijing}
  \country{China}
}

\author{Jincheng An}
\email{anjincheng@qianxin.com}
\affiliation{%
  \institution{Qi An Xin Technology Group Inc.}
  \city{Beijing}
  \country{China}
}

\author{Liehuang Zhu}
\email{liehuangz@bit.edu.cn}
\affiliation{%
  \institution{Beijing Institute of Technology}
  \city{Beijing}
  \country{China}
}

\renewcommand{\shortauthors}{Trovato et al.}

\begin{abstract}
This paper introduces a relaxed noise calibration method to enhance data utility while attaining pufferfish privacy. This work builds on the existing $1$-Wasserstein (Kantorovich) mechanism by alleviating the existing overly strict condition that leads to excessive noise, and proposes a practical mechanism design algorithm as a general solution. We prove that a strict noise reduction by our approach always exists compared to $1$-Wasserstein mechanism for all privacy budgets $\epsilon$ and prior beliefs, and the noise reduction (also represents improvement on data utility) gains increase significantly for low privacy budget situations--which are commonly seen in real-world deployments. We also analyze the variation and optimality of the noise reduction with different prior distributions. Moreover, all the properties of the noise reduction still exist in the worst-case $1$-Wasserstein mechanism we introduced, when the additive noise is largest. We further show that the worst-case $1$-Wasserstein mechanism is equivalent to the $\ell_1$-sensitivity method. Experimental results on three real-world datasets demonstrate $47\%$ to $87\%$ improvement in data utility.

\end{abstract}

\ccsdesc[300]{Security and privacy~Privacy protections}


\keywords{Data Privacy, Pufferfish Privacy, Data Utility}


\maketitle

\section{Introduction}
The proliferation of data-driven applications has led to an unprecedented scale of data collection, sharing, and processing. While these applications drive innovation and economic growth, they also lead to privacy risks, particularly the leakage of sensitive information. Therefore, protecting data privacy has become a critical issue in modern computing systems, including cloud computing~\cite{rtss09,pmoes14,gsgkpw18}, distributed systems~\cite{gzt20_isit,gzt20_tifs,rksc16} and large-scale AI systems~\cite{kylgyo23,lhxtx24,zcwgl24}. For instance, large language model, a landmark development in recent years, remain vulnerable to sensitive data leakage. This highlights the urgency for developing data privacy protection techniques across all domains of data usage.

Differential Privacy (DP), proposed by \cite{dwork06,cfka06}, is considered the gold standard in data privacy preservation against differential attack, where an adversary seeks to infer sensitive information by analyzing the differences in outputs (e.g., query answers) resulting from small changes to the input data, in particular, two databases differ in one entry. 
%
%
To attain DP, query answers are randomized to ensure statistical indistinguishability between a pair of secrets or sensitive attributes. There exist many methods for data privatizations, such as shuffling~\cite{mccj22,csuzz19,lttkcy22}, subsampling~\cite{kk23,zw19,sskg24} and noise addition~\cite{gv15,acgmmtz16,dl09}. As the most common and simplest method, the additive noise mechanism is now widely studied and extensively deployed.

Existing studies on DP typically assume that data records are sampled independently. However, in real-world applications, this assumption is often violated, as records within a dataset frequently exhibit dependencies or correlations--commonly referred to as correlated data. These correlations may arise due to social relationships or shared environmental factors. When such dependencies exist, an adversary can exploit the correlation between sensitive attributes and released data to perform more powerful inference attacks, potentially compromising individual privacy even when standard DP mechanisms are applied. To address this challenge of data security in correlated data, pufferfish privacy, a new privacy framework, is proposed in~\cite{da12,da14}. Pufferfish privacy regards the correlation as prior knowledge of the adversary. To protect the sensitive data, pufferfish privacy enforces statistical distinguishability between two output probability distributions conditioned on a pair of secrets is upper bounded by a given privacy budget $\epsilon$. As elaborated in \cite{km12},  pufferfish privacy is a framework that generalizes differential privacy. However, it comes with new challenges in mechanism design. 


Although many mechanisms like \cite{hmd14,ysn15,zzr11} perform well in specific applications, they rely on some assumptions that are difficult to extend. Independently, \cite{swc17} proposed the first mechanism that can be adopted for general pufferfish privacy setting, called the $\infty$-Wasserstein Mechanism. 
%
%
However, $\infty$-Wasserstein metric is not computable due to the non-convexity of the underlying minimization problem~\cite{cdj08,dl19}. This problem has been pointed out in~\cite{ding22} and solved by a conversion to $1$-Wasserstein (Kantorovich) mechanism, where the optimal transport plan (the minimizer) can be calculated directly by system parameters, i.e., the prior distributions specify the intrinsic correlation between public data and sensitive attributes.  
%
%
On the other hand, R\'{e}nyi pufferfish privacy is proposed in~\cite{pbtb24} by relaxing the original framework via the R{\'e}nyi measure, which is attainable by a general Wasserstein mechanism based on the $W_\infty$ metric. Here, the computation problem for $W_\infty$ still exists. 
Additionally, many recent works focus on quantum mechanisms~\cite{ngw24,nsw25,yyc24} and the approximate pufferfish privacy under the assumption that the adversary's prior belief of the published data is Gaussian distributed~\cite{ding24}. These papers further enhance the development of pufferfish privacy.

Prior research has mainly focused on the design of generic mechanisms, computational problems, and scalability. 
While these efforts have advanced the practical deployment of the pufferfish privacy frameworks, less attention has been paid to optimizing the data utility. 
Specifically, the $1$-Wasserstein mechanism~\cite{ding22} adjusts the noise parameter to the maximum pairwise distance over Kantorovich optimal transport plan. But, such a transport plan is itself a probability distribution, where the maximum pairwise distance could be least likely to appear in reality. In this case, $1$-Wasserstein mechanism may generate too much noise that unnecessarily degrades data utility. 

In this work, we shift the perspective to noise minimization--a critical yet underexplored area in the design of privacy-preserving mechanisms. 
Our study focuses on the $1$-Wasserstein (Kantorovich) mechanism~\cite{ding22}. We reveal that $1$-Wasserstein mechanism adds excessive noise to the data. The reason is that it enforces a strict sufficient condition for attaining pufferfish privacy. 
We summarize our main contributions as follows.
\begin{itemize}
    \item We derive a relaxed sufficient condition that will generate a smaller noise parameter than the $1$-Wasserstein mechanism. We propose a practical mechanism design algorithm to search for such a noise parameter. This algorithm applies to any pufferfish privacy setting with finite and countable alphabet. 
    \item We prove that there is always a noise reduction if replacing the $1$-Wasserstein mechanism by our proposed relaxed mechanism design, for any privacy budget $\epsilon$. In particular, in the low privacy regime $\epsilon \in (0,1]$, the noise reduction is significantly large, indicating a great enhancement in data utility. Experimental results on three real-world datasets demonstrate \textbf{ 47\% to 87\% higher data utility}, validating the practical impact of our approach.
    \item Considering the worst-case $1$-Wasserstein mechanism, an extreme case when this mechanism generates the largest noise that severely deteriorates data utility, we show that it is equivalent to applying the $\ell_1$-sensitivity noise calibration method, and a meaningful noise reduction can be achieved by adopting our approach to maintain the usefulness of released data. 
\end{itemize}

This paper presents a self-contained and practical noise calibration method for pufferfish privacy, along with a theoretical analysis of its noise reduction.
The paper is organized as follows. 
Section~\ref{sec:prelim} reviews standard definitions and existing mechanisms for achieving pufferfish privacy. 
Section~\ref{sec:main_method} outlines the motivation behind this work and introduces a practical relaxed mechanism, including its implementation details and theoretical derivations. 
Section~\ref{sec:noise_reduction} demonstrates that our proposed mechanism strictly reduces noise across all privacy budgets, even in the worst-case 1-Wasserstein mechanism analyzed in Section~\ref{sec:w1_equal_dp}. 
Section~\ref{sec:experiment} evaluates the performance of our approach through real-world experiments. 
Finally, Section~\ref{sec:conclusion} concludes with future directions.

\section{Preliminaries}\label{sec:prelim}
We review the pufferfish privacy framework and the noise calibration methods by Wasserstein metric.

{\bf Pufferfish privacy.} Let $S$ be sensitive attribute and $X$ be the data we want to publish. $S$ is assumed to be correlated with $X$ and therefore a direct publication reveals sensitive information on $S$, i.e., a breach of data privacy. For example, publishing the heart disease will reveal the individual's age group. Let $\rho$ be the prior knowledge of the adversary concerning the correlation $P_{X|S}(\cdot|s,\rho)$ for all $s$. For example, $P_{X|S}(\cdot|s,\rho)$ could be a Gaussian distribution for a counting query with the mean and covariance inferred from previous data releases. There could be more than one adversary in the system, and each $\rho$ uniquely identifies one of them. We will use the notation $\max_{\rho}$ in the main context, which refers to the maximization over all adversaries. The interpretation of this maximization is to provide privacy protection against all adversaries.

Let $\mathbb{S}$ be a subset of secret pairs $(s_i,s_j)$. The adversary is assumed to have access to the privatized data $Y$ only. He can collect the aggregated statistics by repeatedly querying the released database. To protect privacy, the randomized mechanism should be carefully designed to ensure enough statistical indistinguishability between all secret pairs $(s_i,s_j)\in \mathbb{S}$. For $\epsilon>0$ being the \emph{privacy budget}, $Y$ is called ($\epsilon$,$\mathbb{S}$)-\emph{pufferfish privacy} if \cite{da14}
    \begin{align}\label{eq:pufferfish_privacy}
        e^{-\epsilon} \leq \frac{P_{Y|S}(y|s_i,\rho)}{P_{Y|S}(y|s_j,\rho)} \leq e^{\epsilon}, \quad \forall (s_i,s_j) \in \mathbb{S}, \rho.
    \end{align} 



{\bf Additive noise mechanism and data utility}. \label{sec:additive_noise_scheme}
To protect privacy, we add independent noise $N$ to $X$ and release the randomized data $Y = X + N$. Denote $P_N(\cdot)$ as the probability of $N$. Then, $P_{Y|S}(y|s,\rho)$ can be computed by a convolution
\begin{align}\label{eq:prob_y_given_s}
    P_{Y|S}(y|s,\rho) = \int P_N(y-x) P_{X|S}(x|s,\rho)\D x.
\end{align}
The noise variance $\mathrm{VAR}[N] = \mathbb{E}[(Y-X)^2]$ represents the mean squared error~(MSE) between the original and randomized data, indicating the loss in data utility~\cite{hmd14}. 
For \emph{Laplace noise} $N_{\theta}$~($\theta$ is the parameter of Laplace noise), the probability density function is $P_{N_\theta}(z) = \frac{1}{2\theta}e^{-\frac{|z|}{\theta}}$ and the variance is $\mathrm{VAR}[N_{\theta}] = 2\theta^2$, i.e.,
\begin{equation} \label{eq:VAR}
    \mathrm{MSE} = \mathbb{E}[(Y-X)^2] = E[N^2] = \mathrm{VAR}[N] = 2\theta^2.
\end{equation}
Therefore, a smaller $\theta$ will produce less noise and preserve data utility. In this work, we use $\theta$ to measure data utility in simulations (Section~\ref{sec:noise_reduction}, Section~\ref{sec:worst_case_w_1}) and experiments (Section~\ref{sec:experiment}).

{\bf Noise calibration by Wasserstein metric. }For each prior belief $\rho$, a joint distribution $\pi:\mathbb{R}^2 \mapsto [0,1] $ is called a \emph{coupling} of $P_{X|S}(\cdot|s_i,\rho)$ and $P_{X|S}(\cdot|s_j,\rho)$ if they are two marginals of $\pi$, i.e., $P_{X|S}(x|s_i,\rho) = \int \pi(x,x') \D x'$ for all $x$ and $P_{X|S}(x'|s_j,\rho) = \int \pi(x,x') \D x$ for all $x'$. 
The $1$-Wasserstein distance is defined as $W_1(s_i,s_j) := \inf_{\pi}\int |x-x'| \D \pi(x,x^{\prime})$ which corresponds to the Kantorovich optimal transport problem~\cite{kantorovich06}.
The minimizer is the Kantorovich optimal transport plan, denoted by $\pi^*$. 
A \emph{$W_1$ (Kantorovich) mechanism} is proposed in~\cite[Lemma~1]{ding22} stating that adding Laplace noise $N_{\theta_1}$ with
\begin{align}\label{eq:w_1_mechanism}
    \theta_1 = \frac{1}{\epsilon}\max_{\rho, (s_i, s_j) \in \mathbb{S}} \sup_{(x,x') \in \supp(\pi^*)} |x-x'|
\end{align}
guarantees $(\epsilon, \mathbb{S})$-pufferfish privacy, where $\supp(\pi^*)$ stands for the support of $\pi^*$. It is shown that the $W_1$ mechanism is equivalent to the $W_\infty$ mechanism in~\cite{swc17}, but much easier to compute. Without solving the minimization problem, $\pi^*$ can be directly determined from $P_{X|S}(\cdot|s_i,\rho)$ and $P_{X|S}(\cdot|s_j,\rho)$.~\cite{ding22}

{\bf Limitation.} The $W_1$ mechanism in \cite[Lemma~1]{ding22} imposes a strict pointwise constraint by requiring $\frac{ |x-x'| }{ \theta } \leq \epsilon$ for all $(x,x') \in \supp( \pi^* )$. However, in the derivation of pufferfish privacy introduced in Eq.~\eqref{eq:triangleIneq}, the term $(e^{\frac{|x-x'|}{\theta}} - e^{\epsilon})$ is first averaged over the coupling $\pi^*(x,x')$ and then further smoothed by the noise kernel $P_{N_{\theta}}(y-x')$. Due to this inherent randomization, the pointwise requirement that each $(e^{\frac{|x-x'|}{\theta}} - e^{\epsilon})$ be non-positive is overly strict and may lead to a large noise parameter $\theta$. Consequently, such a large $\theta$ may introduce excessive additive noise, unnecessarily decreasing data utility to satisfy statistical indistinguishability. To address this limitation, we provide a theoretical analysis in Section~\ref{sec:strict_relaxed_condition}, relaxing the overly strict condition to better align with the expected-distance formulation and achieve privacy-utility trade-off.

\section{Practical Noise Calibration Mechanism}\label{sec:main_method}
\begin{figure*}[!ht]
    \centering
    \subfloat[Strict condition]{%
        \includegraphics[width=0.5\textwidth]{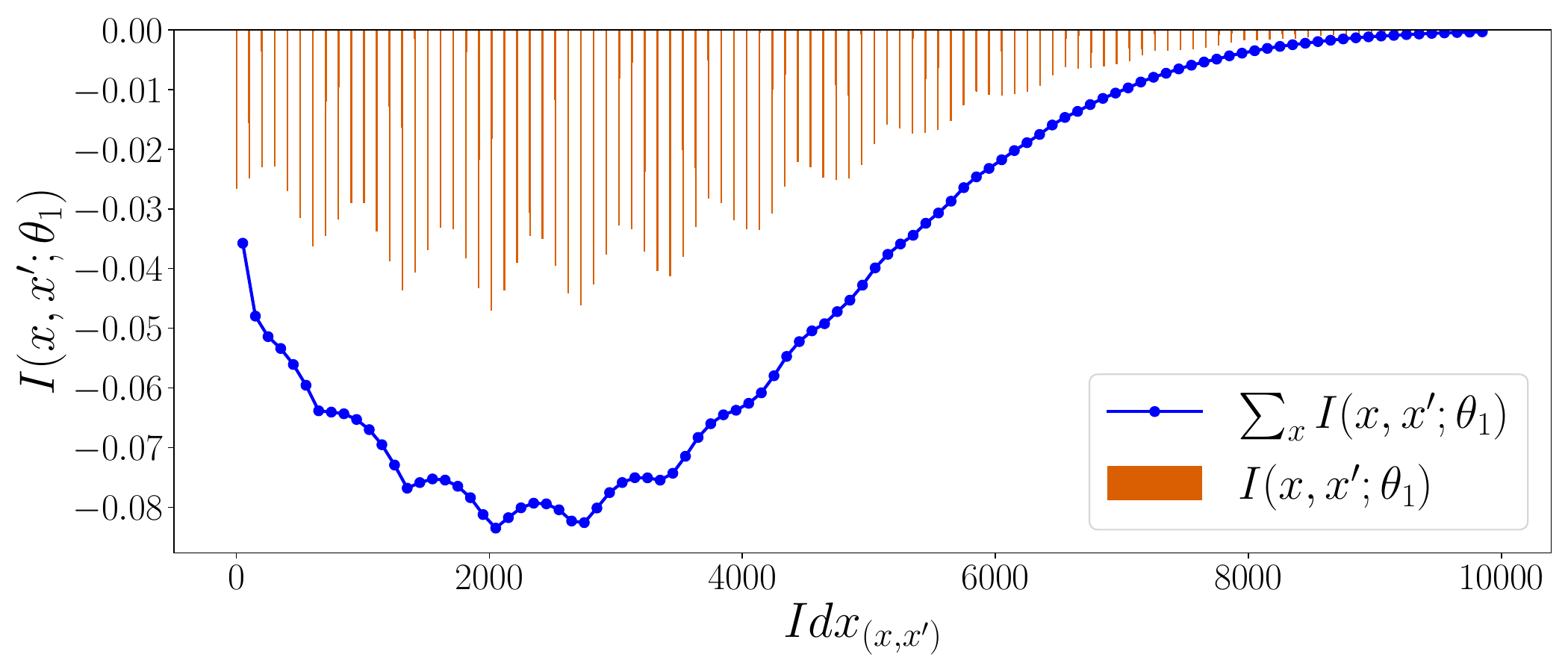}
        \label{fig:relaxed_source_strict}
    }
    \subfloat[Relaxed condition]{%
        \includegraphics[width=0.5\textwidth]{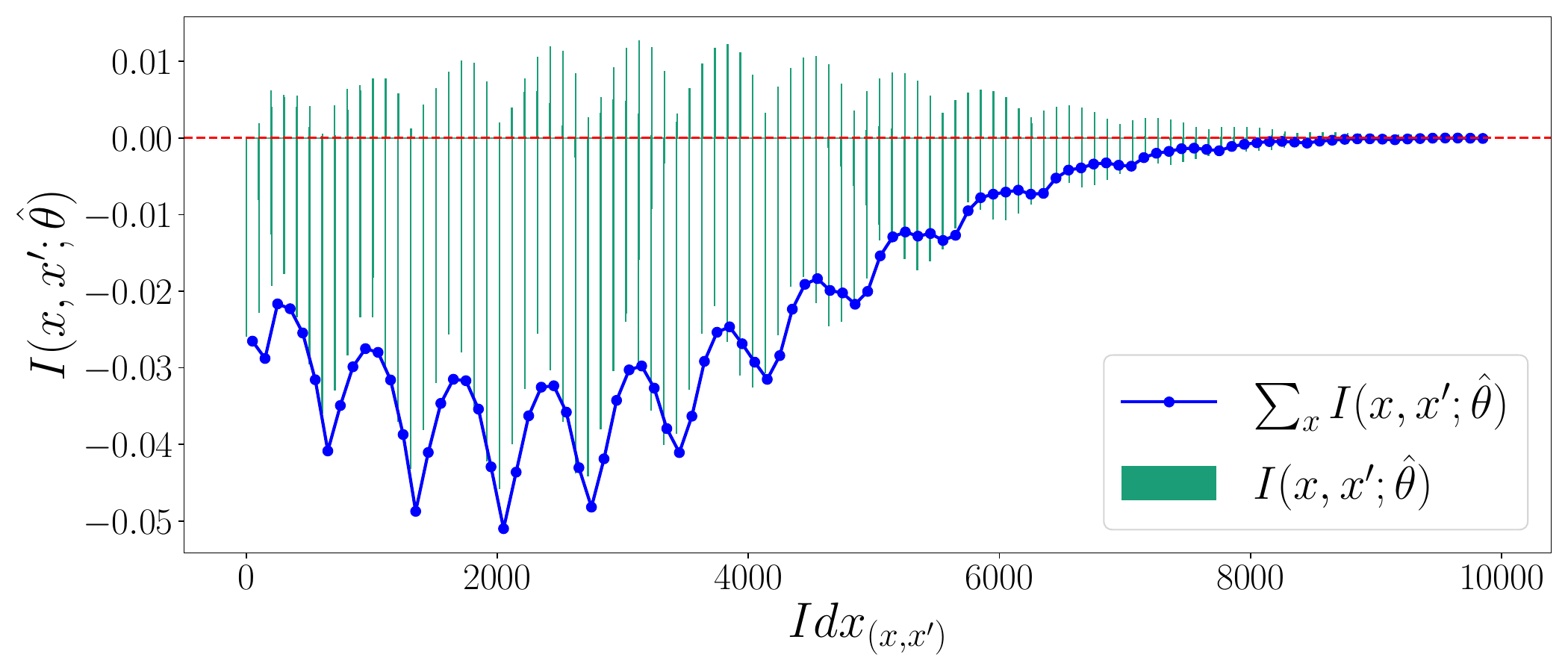}
        \label{fig:relaxed_source_relaxed}
    }
    \caption{Visualization of strict and relaxed condition: \ref{fig:relaxed_source_strict} shows that $I(x,x';\theta_1)\leq 0$ holds for all $(x,x')\in\mathcal{X}^2$ satisfying Eq.~\eqref{eq:w_1_mechanism}. Consequently, $\sum_x I(x,x';\theta_1) \leq 0$ for all $x'\in\mathcal{X}$ satisfying $(\epsilon,\mathbb{S})$-pufferfish privacy. \ref{fig:relaxed_source_relaxed} shows that while $\sum_x I(x,x';\theta_1) \leq 0$ still holds for all $x'\in\mathcal{X}$ satisfying pufferfish privacy, not all $I(x,x';\theta_1)$ are non-positive--the relaxation of the strict condition. The horizontal axis indexes the ordered pairs $(x,x')$, sorted primarily by $x'$ and then by $x$. The index starts at $1$ for $(0,0)$, increases sequentially with $x$, and wraps to the next value of $x'$ after every $100$ steps. For instance, $(99,0)$ has index $100$ and $(0,1)$ has index $101$.}
\end{figure*}

In this section, we first introduce the derivation of the strict condition and explain its limitations. 
Then, we introduce the relaxed mechanism, which is motivated by the limitations of the $W_1$ mechanism. Then, we propose a practical noise calibration method along with the algorithmic implementation.

\subsection{Motivation: from Strict to Relaxed Sufficient Condition}\label{sec:strict_relaxed_condition}
Following the definition of pufferfish privacy \eqref{eq:pufferfish_privacy}, it is clear that $(\epsilon, \mathbb{S})$-pufferfish privacy attains at secret pair $(s_i,s_j)$ if $P_{Y|S}(y|s_i,\rho) - e^{\epsilon}P_{Y|S}(y|s_j,\rho) \leq 0$ and $P_{Y|S}(y|s_j,\rho) - e^{\epsilon}P_{Y|S}(y|s_i,\rho) \leq 0$ for all $y$ and $\rho$. 
By the convolution~\eqref{eq:prob_y_given_s}, we can work out an upper bound on the left hand side, assuming Laplace noise $N_\theta$ is added to $X$:\footnote{Due to symmetry, we only analyze $P_{Y|S}(y|s_i,\rho) - e^{\epsilon}P_{Y|S}(y|s_j,\rho) \leq 0$ for a prior $\rho$. The counterpart, $P_{Y|S}(y|s_j,\rho) - e^{\epsilon}P_{Y|S}(y|s_i,\rho) \leq 0$, follows similarly. Eq~\eqref{eq:start} to Eq~\eqref{eq:triangleIneq} are shown in~\cite{ding22}. We restate them for the self-containment of this paper. } 

\begin{align}
    P_{Y|S}(& y|s_i,\rho)-e^{\epsilon}P_{Y|S}(y|s_j,\rho) \notag\\
    & = \int(P_{N_{\theta}}(y-x)-e^{\epsilon}P_{N_{\theta}}(y-x'))\D \pi^*(x,x'), \label{eq:start} \\
    & = \int \frac{1}{2\theta} \Big( e^{-\frac{|y-x|}{\theta}} - e^{\epsilon-\frac{|y-x'|}{\theta}} \Big) \D \pi^*(x,x'), \notag \\
    & = \int \frac{1}{2\theta} e^{-\frac{|y-x'|}{\theta}}  \Big( e^{\frac{|y-x'|-|y-x|}{\theta}} - e^{\epsilon} \Big)  \D \pi^*(x,x'), \notag\\
    & \leq \int P_{N_{\theta}}(y-x')\underbrace{\Big(e^{\frac{|x-x'|}{\theta}} - e^{\epsilon} \Big)}_{\leq 0, W_1 \text{ mechanism}} \D \pi^*(x, x'),\label{eq:triangleIneq}\\
    & = \int P_{N_{\theta}}(y-x') \underbrace{ \Big( \int (e^{\frac{|x-x'|}{\theta}} - e^{\epsilon}) \pi^*(x,x')\D x  \Big) }_{\leq 0, \text{ relaxed condition}} \D x'. \label{eq:relaxed_condition_source}
\end{align}

Knowing that $P_{N_{\theta}}(z) \in [0,1]$ for all $z$ and $\pi^*(x,x')$ is non-negative for all $x$ and $x'$, there are two approaches to have $P_{Y|S}(y|s_i,\rho)-e^{\epsilon}P_{Y|S}(y|s_j,\rho) \leq 0$ for all $y$.
One is to request 
\begin{equation}\label{eq:Strict}
    e^{\frac{|x-x'|}{\theta}} - e^{\epsilon} \leq 0, \quad \forall (x,x') \in \supp(\pi^*)
\end{equation}
in Eq.~\eqref{eq:triangleIneq}. That is, 
\begin{align*}
    \sup_{(x,x') \in \supp(\pi^*)} e^{\frac{|x-x'|}{\theta}} - e^\epsilon = 0.   
\end{align*}
This gives the $W_1$ mechanism~(Eq.~\eqref{eq:w_1_mechanism}).
The other is to have the inner integral in Eq.~\eqref{eq:relaxed_condition_source} non-positive, i.e., 
\begin{equation} \label{eq:Relaxed1}
\int (e^{\frac{|x-x'|}{\theta}} - e^{\epsilon}) \pi^*(x,x')\D x \leq 0, \quad \forall x'
\end{equation}
As maximum is no greater than the expectation, it is clear that Eq.~\eqref{eq:Relaxed1} is a sufficient condition relaxed from Eq.~\eqref{eq:Strict}, which can produce a smaller $\theta$ for attaining $(\epsilon,\mathbb{S})$-pufferfish privacy. 
Specifically, define
\begin{align*}
    I(x,x';\theta) := (e^{\frac{|x-x'|}{\theta}} - e^{\epsilon}) \pi^*(x,x').
\end{align*} 
The smallest value of $\theta$ that satisfies the relaxed sufficient condition Eq.~\eqref{eq:Relaxed1} is the one that holds the following equations
\begin{equation} \label{eq:RelaxedPair}
    \begin{aligned}
        & \int I(x,x';\theta) \D x = 0,\quad \forall x', \\
        & \int I(x,x';\theta) \D x' = 0, \quad \forall x. 
    \end{aligned}
\end{equation}
This involves solving integral equations. 

It should be noted that Eq.~\eqref{eq:RelaxedPair} has also been proposed in~\cite[Theorem~2]{ding22}. 
However, \cite{ding22} only states that there exists a smaller value of $\theta$ by relaxing the sufficient condition for the $W_1$ mechanism. 
Although~\cite{ding22} mentioned that the noise parameter can be determined by solving a polynomial equation, but does not provide a concrete noise calibration mechanism specifying how to set the exact value of the noise parameter $\theta$, e.g., a closed-form expression of the $\ell_1$-sensitivity method similar to \cite{dmns06} for differential privacy. In fact, the difficulty is how to solve high-order polynomial equations\footnote{It is stated in Abel–Ruffini Theorem~\cite{ayoub80} that there is no solution in radicals to general polynomial equations of degree more than 5 with arbitrary coefficients.}, while \cite{ding22} avoided this difficulty by only solving polynomial equations of degree 1, which resulted from the restricted pairwise distance in Kantorovich optimal transport plan $\max_{(x,x') \in\supp(\pi^*)}|x-x'| \leq 1$. However, this is not the case in general.

\begin{assumption}[Discrete Setting]
    Let the support of $\pi^*(x,x')$ be a countable and finite alphabet. This is usually the case in practice. In this case, integral in Eq.~\eqref{eq:relaxed_condition_source} reduces to summation: 
    \begin{align*}
        \sum_{x'} P_{N_{\theta}}(y-x') \sum_x I(x,x';\theta).
    \end{align*}
\end{assumption}

To visualize this relaxation, we calculate both the strict and the relaxed conditions in the medical dataset~\cite{diabetes} as an example. 
See an experimental example in the medical dataset we treat `Glucose' as \emph{public attribute} $X$, which will be released (e.g., for statistical analysis), whereas `Age' is designated as \emph{sensitive attribute} $S$,  which should be protected under privacy constraints. 
We discretize the range of `Glucose' into 100 levels, i.e., $\supp(\pi^*)$ is discrete with $x, x' = 0, \ldots, 99$. 
Specifically, in this example with privacy budget $\epsilon=1$, we compute $\theta_1$ under the $W_1$ mechanism (Eq.~\eqref{eq:w_1_mechanism}) and plot in Figure~\ref{fig:relaxed_source_strict}. 

The bars in this figure represent the values of $I(x,x';\theta_1)$ for all $(x,x')\in\supp(\pi^*)$ and the dotted plot represent the values of $\sum_x I(x,x';\theta_1)$ for all $x'$. 
Additionally, with $\epsilon=1$, we also calculate $\hat{\theta}$ following the practical noise calibration method (Introduced in Section~\ref{sec:approximate_results}) and show it in Figure~\ref{fig:relaxed_source_relaxed}. 
The experimental visualization in both Figure~\ref{fig:relaxed_source_strict} and Figure~\ref{fig:relaxed_source_relaxed} illustrates three points. 
\begin{enumerate}
    \item In these figures, $\sum_x I(x,x';\theta_1) \leq 0$ and $\sum_x I(x,x';\hat{\theta}) \leq 0$ for all $x'$. This means setting the Laplace noise parameter to either $\theta_1$ or $\hat{\theta}$ is sufficient to attain $(\epsilon,\mathbb{S})$-pufferfish privacy.
    \item In Figure~\ref{fig:relaxed_source_strict}, for all pairs of $(x,x')$, the value of $I(x,x';\theta_1)$ is negative, and then the summation $\sum_x I(x,x';\theta_1)$ for all $x'$ is also negative.
    \item In Figure~\ref{fig:relaxed_source_relaxed}, the summation $\sum_x I(x,x';\hat{\theta})$ for all $x'$ is non-positive while not all the $I(x,x';\hat{\theta})$ are negative.
\end{enumerate}
These findings indicate that both strict conditions and relaxed conditions can attain the same level of pufferfish privacy. However, the condition~\eqref{eq:w_1_mechanism} in the $W_1$ mechanism is overly strict, as not all the $I(x,x';\theta_1)$ needs to be negative. This necessarily leads to more additive noise while attaining $(\epsilon,\mathbb{S})$-pufferfish privacy. We will explain and prove this is the case in Section~\ref{sec:noise_reduction}.

Following~\eqref{eq:RelaxedPair}, we relax the sufficient condition \eqref{eq:w_1_mechanism} supports $W_1$ mechanism to 
\begin{align}
    & \sum_x I(x, x'; \theta) \leq 0, \quad \forall x',\label{eq:discrete_relaxed_condition_x_prime}\\
    & \sum_{x'} I(x, x'; \theta) \leq 0, \quad \forall x, \label{eq:discrete_relaxed_condition_x}
\end{align}
We only consider \eqref{eq:discrete_relaxed_condition_x_prime} in this paper, as the other can be derived in the same way. 

\subsection{Relaxed Mechanism}\label{sec:relaxation_method}

While~\eqref{eq:discrete_relaxed_condition_x_prime} only states the condition that an $(\epsilon,\mathbb{S})$-pufferfish privacy attaining $\theta$ should satisfy, the following question is how to determine such a $\theta$, i.e., 

{\bf Q1.} \textit{How to apply the relaxed conditions~\eqref{eq:discrete_relaxed_condition_x_prime} to calibrate the noise parameter $\theta$?} 

To answer this question, we propose a noise calibration mechanism below.
Define the LHS of \eqref{eq:discrete_relaxed_condition_x_prime} by
\begin{align*}
  f_{x'}(\theta) := \sum_x I(x, x'; \theta).
\end{align*}
Thus, the problem of searching for a minimum value of $\theta$ satisfying \eqref{eq:discrete_relaxed_condition_x_prime} reduces to finding the root of $f_{x'}(\theta)=0$.  
To simplify, we denote $\hat{\theta}$ as the root of $f_{x'}(\theta)=0$ and $\hat{\theta}$ is also presented as $f_{x'}^{-1}(0)$. 
Then, we propose the relaxed mechanism as follows.
\begin{proposition}[Relaxed Mechanism]\label{pro:relaxed_condition_mechanism}
    Adding Laplace noise $N_{\hat{\theta}}$ with
    \begin{equation}\label{eq:Relaxed}
        \hat{\theta} = \max_{\rho\in\mathbb{D},(s_i,s_j)\in\mathbb{S}} \sup_{(x,x')\in\supp(\pi^*)} f_{x'}^{-1}(0)
    \end{equation}
    attains $(\epsilon, \mathbb{S})$-pufferfish privacy in $Y$.
\end{proposition}
\begin{proof}
    From the relaxed condition \eqref{eq:discrete_relaxed_condition_x_prime}, we have
    \begin{align*}
        f_{x'}(\theta) 
        & = \sum_x I(x, x'; \theta) =\sum_x \big( e^\frac{|x-x'|}{\theta}-e^{\epsilon} \big)\pi^*(x,x'),
    \end{align*}
    As $\pi^*(x,x')$ are non-negative for all pairs of $(x,x')$, $f_{x'}(\theta)$ is decreasing functions of $\theta$. Thus, the root $\hat{\theta}$ in Eq.~\eqref{eq:Relaxed} satisfies $f_{x'}(\hat{\theta}) \leq 0$ for all $x'\in\mathcal{X}$. 
    The relaxed mechanism in this proposition can attain $(\epsilon, \mathbb{S})$-pufferfish privacy.
\end{proof}

\subsection{Practical Relaxed Mechanism}\label{sec:approximate_results}
As mentioned above, the problem of searching for $\hat{\theta}$ in Proposition~\ref{pro:relaxed_condition_mechanism} is a root-finding problem of $f_{x'}(\theta)=0$. As 
\begin{align*}
    f_{x'}(\theta) 
    & = \sum_x \big( e^\frac{|x-x'|}{\theta}-e^{\epsilon} \big)\pi^*(x,x')\\
    & = \sum_x \Big( \big(e^\frac{1}{\theta}\big)^{|x-x'|}-e^{\epsilon} \big)\pi^*(x,x')
\end{align*}
is a polynomial of $e^\frac{1}{\theta}$ and the order of this polynomial is determined by $n := \max_{(x,x')\in\supp(\pi^*)}|x-x'|$ (which we denote as $n$). Due to the Abel–Ruffini Theorem~\cite{ayoub80}, if $n \geq 5$, $\hat{\theta}$ is not expressible in radicals, which raises a question for the usage of the relaxed mechanism in practice.

{\bf Q2.} \textit{How to determine noise parameter $\hat{\theta}$ for arbitrary polynomial order $n$, especially high-order polynomials?}

In numerical analysis, Brent's method~\cite{brent13}, a root-finding algorithm, is often regarded as a reliable approach for approximating the roots of high-order polynomial equations. 
Brent's method begins with two initial points $a$ and $b$ such that $f(a)\cdot f(b) < 0$, ensuring that the root is local within the interval $[a,b]$. 
At each iteration, the algorithm selects the next estimate to make $a$ and $b$ become closer until the width of the interval falls below a specified tolerance. Then, the bound in the final iteration, $a'$ or $b'$, is regarded as the approximate root of the polynomial $f$.

To determine this initial interval in our mechanism, we first theoretically derive two initial points $\hat{\theta}_a$ and $\hat{\theta}_b$ that bound the noise parameter $\hat{\theta}$ in all cases, i.e., $\hat{\theta} \in [\hat{\theta}_a, \hat{\theta}_b]$. 
Let
\begin{align}
    \Phi := \ln\frac{e^{\epsilon} P(x'|s_j, \rho) - \pi^{*}(x',x')}{{P(x'|s_j, \rho)-\pi^{*}(x', x')}}, \label{eq:initial_para_brent_method}
\end{align} 
denote the initial parameter (Algorithm~\ref{alg:brent_method}, Line~\ref{alg:line_initial_parameter}) and $n = \\\max_{(x,x')\in\supp(\pi^*)}|x-x'|$ denote the order of the polynomial equation $f_{x'}(\theta)$. 
We set the the initial points $\hat{\theta}_a=1/\Phi$ and $\hat{\theta}_b=n/\Phi$, which satisfy $f_{x'}(\hat{\theta}_a)>0$ and $f_{x'}(\hat{\theta}_b)<0$ (Algorithm~\ref{alg:brent_method}, Line~\ref{alg:line_initial_points}). 
These two initial points achieve $f_{x'}(\hat{\theta}_a) \cdot f_{x'}(\hat{\theta}_b) \leq 0$ thus $\hat{\theta}\in [\hat{\theta}_a,\hat{\theta}_b]$. It guarantees that the root is enclosed within the interval lower bounded by $\hat{\theta}_a$ and upper bounded by $\hat{\theta}_b$, enabling reliable convergence.

As the algorithm runs, $\hat{\theta}_a$ and $\hat{\theta}_b$ becomes closer until $\hat{\theta}'_b - 
\hat{\theta}'_a \leq \nu$, where $\nu$ is the error tolerance, $[\hat{\theta}'_a, \hat{\theta}'_b]$ are the interval of the final iteration. 
The original Brent's method returns one of these two approximate results as the approximate root. However, as $f_{x'}(\hat{\theta}'_a) > 0$, the smaller $\hat{\theta}'_a$ does not satisfy condition~\eqref{eq:discrete_relaxed_condition_x_prime} to attain $(\epsilon, \mathbb{S})$-pufferfish privacy. 
To address this, we modify the algorithm to return the larger $\hat{\theta}'_b$ as the approximate root (Algorithm~\ref{alg:brent_method}, Line~\ref{alg:line_max_results}), which satisfies $f_{x'}(\hat{\theta}'_b)<0$ to attain $(\epsilon, \mathbb{S})$-pufferfish privacy (Eq.~\eqref{eq:discrete_relaxed_condition_x_prime}). 

After the modifications above, we can calculate the approximate root of $f_{x'}$. We run modified Brent's method for all $x' \in \mathcal{X}$ and choose the maximum value as the noise parameter $\hat{\theta}$.  
Then we propose a practical relaxed mechanism (Proposition~\ref{pro:practical_relaxed_mechanism}) and provide its algorithmic implementation (Algorithm~\ref{alg:brent_method}) as follows. 

\begin{algorithm}[t]
\caption{Practical Relaxed Mechanism}
\label{alg:brent_method}
\KwIn{Probability distributions $P_{X|S}(x|s_i, \rho)$, $P_{X|S}(x'|s_j, \rho)$, Kantorovich optimal transport plan $\pi^*(x,x')$, privacy budget $\epsilon$, error tolerance $\nu$}
\KwOut{Approximate noise parameter $\hat{\theta}$}
\BlankLine
\Comment{Initialize candidate set $C$}
$C \gets \emptyset$\;

\Comment{Calculate $\hat{\theta}$ of $f_{x'}(\theta)$ for each $x'$}
\ForEach{$x' \in \mathcal{X}$}{
    \Comment{Define polynomial function}
    $f_{x'}(t) = \sum_x \big( t^{|x-x'|}-e^{\epsilon} \big)\pi^*(x,x')$
    
    \Comment{Initialize parameter from Eq.~\eqref{eq:initial_para_brent_method}}
    $\Phi \gets \ln\!\Big(\frac{\,e^{\epsilon} P(x'|s_j,~\rho) - \pi^*(x',x')\,}{\,P(x'|s_j,~\rho) - \pi^*(x',x')\,}\Big)$\;\label{alg:line_initial_parameter}
    
    \Comment{Set initial points from Proposition~\ref{pro:practical_relaxed_mechanism}~(\ref{pro:practical_relaxed_mechanism_1})}
    $\hat{\theta}_a \gets 1/\Phi$, $\hat{\theta}_b \gets N/\Phi$\;\label{alg:line_initial_points}
    
    \Comment{Transform to exponential form}
    $t_a \gets e^{1/\hat{\theta}_a}$, $t_b \gets e^{1/\hat{\theta}_b}$
    
    \Comment{Apply Brent's method}
    \lIf{$|f_{x'}(t_b)| > |f_{x'}(t_a)|$}{
        Swap $t_a$ and $t_b$
    }
    
    $t_c \gets t_a$\;
    
    \While{$|t_b - t_c| > \nu$ \textbf{and} not max iterations}{
    
        \uIf{$f_{x'}(t_a) \neq f_{x'}(t_c)$ \textbf{and} $f_{x'}(t_b) \neq f_{x'}(t_c)$}{
            \Comment{Inverse quadratic interpolation}
            $t_{b}' \gets \text{IQI}(t_a, t_b, t_c)$\;
        }
        \uElseIf{$f_{x'}(t_a) \neq f_{x'}(t_b)$}{
            \Comment{Secant step}
            $t_{b}' \gets \frac{t_a f_{x'}(t_b) - t_b f_{x'}(t_a)}{f_{x'}(t_b) - f_{x'}(t_a)}$\;
        }
        \Else{
            \Comment{Bisection fallback}
            $t_b' \gets (t_b + t_c)/2$\;
        }

        \If{$t_b'$ not in $(t_b, t_c)$}{
            $t_b' \gets (t_b + t_c)/2$\;
        }
        \Comment{Update interval}
        $t_a \gets t_c$, $t_c \gets t_b$, $t_b \gets t_b'$\;
    }
    
    \Comment{Transform back to $\hat{\theta}$ domain}
    $\hat{\theta}'_a \gets 1/\ln(t_c)$, $\hat{\theta}'_b \gets 1/\ln(t_b)$
    
    \Comment{Add candidate result}
    $C$.append$\Big( \max(\hat{\theta}'_a ,\hat{\theta}'_b) \Big)$\label{alg:line_max_results}
}
\Comment{Final output: maximum over all candidates}
$\hat{\theta} \gets \max (C)$\;
\Return $\hat{\theta}$

\end{algorithm}

\begin{proposition}(Practical Relaxed Mechanism)\label{pro:practical_relaxed_mechanism}
    Adding Laplace noise $N_{\hat{\theta}}$ attains $(\epsilon, \mathbb{S})$-pufferfish privacy, where
    \begin{enumerate}
        \item $\hat{\theta} \in [\hat{\theta}_a,\hat{\theta}_b]$ always exists for any $\pi^*(x,x')$ and $\epsilon$ in each iteration of Algorithm~\ref{alg:brent_method},\label{pro:practical_relaxed_mechanism_1}
        \item $\hat{\theta}$ can be approximated by Algorithm~\ref{alg:brent_method}.\label{pro:practical_relaxed_mechanism_2}
    \end{enumerate}
\end{proposition}
\begin{proof}
    For $f_{x'}(\hat{\theta})=0$, we define two points $\hat{\theta}_a$ and $\hat{\theta}_b$ such that $f_{x'}(\hat{\theta}_a) > 0$ and $f_{x'}(\hat{\theta}_b) < 0$. To make $f_{x'}(\hat{\theta}_a)>0$, 
    
    \begin{align*}
      f_{x'}(\hat{\theta}_a) &= \sum_x e^{\frac{|x-x'|}{\hat{\theta}_a}} \pi^{*}(x,x') - e^{\epsilon}P(x'|s_j, \rho) \notag\\
      &\geq e^{\frac{1}{\hat{\theta}_a}} \sum_{x \neq x'} \pi^{*}(x,x') + \pi^{*}(x', x') - e^{\epsilon}P(x'|s_j, \rho) 
    \end{align*}
    where the third step follows from $\sum_{i=1}^n x^i \geq n \cdot x$ when $x>1$.
    
    Then we have
    \begin{align*}
      e^{\frac{1}{\hat{\theta}_a}} (P(x'|s_j, \rho)-\pi^{*}(x', x')) > e^{\epsilon}P(x'|s_j, \rho) - \pi^{*}(x',x')
    \end{align*}
    Then we get
    \begin{align*}
      \hat{\theta}_a = \frac{1}{\ln(e^{\epsilon} P(x'|s_j, \rho) - \pi^{*}(x',x'))-\ln({P(x'|s_j, \rho)-\pi^{*}(x', x')})}  
    \end{align*}
    such that $f_{x'}(\hat{\theta}_a)>0$. 
    To make $f_{x'}(\hat{\theta}_b) < 0$,
    \begin{align*}
      f(\hat{\theta}_b) &= \sum_x e^{\frac{|x-x'|}{\hat{\theta}_b}} \pi^{*}(x,x') - e^{\epsilon}P(x'|s_j, \rho) \\
      &\leq e^{\frac{n}{\hat{\theta}_b}} \sum_{x \neq x'} \pi^{*}(x,x') + \pi^{*}(x', x') - e^{\epsilon}P(x'|s_j, \rho)
    \end{align*}
    where the third step follows from $\sum_{i=1}^n x^i \leq n \cdot x^n$ when $x>1$
    Then we have
    \begin{align*}
      e^{\frac{n}{\hat{\theta}_b}} (P(x'|s_j, \rho)-\pi^{*}(x', x')) < e^{\epsilon}P(x'|s_j, \rho) - \pi^{*}(x',x')
    \end{align*}
    We get another init point 
    \begin{align*}
      \hat{\theta}_b = \frac{n}{ \ln(e^{\epsilon} P(x'|s_j, \rho) - \pi^{*}(x',x'))-\ln({P(x'|s_j, \rho)-\pi^{*}(x', x')})  }
    \end{align*}
    such that $f_{x'}(\hat{\theta}_b)<0$. 
    
    Define the support of $\pi^*$ into two space as $A=\{(x,x') \in \supp(\pi^*):|x-x'|\geq 1\}$ and $B=\{(x,x') \in \supp(\pi^*):x=x'\}$, then we denote
    \begin{align}
        \Phi 
        & := \ln(e^{\epsilon} P(x'|s_j, \rho) - \pi^{*}(x',x'))-\ln({P(x'|s_j, \rho)-\pi^{*}(x', x')}), \notag\\
        & = \ln\Big( \frac{e^{\epsilon}(\sum_A\pi^*(x,x')+\sum_B\pi^*(x,x')) - \sum_B\pi^*(x,x')}{(\sum_A\pi^*(x,x')+\sum_B\pi^*(x,x'))-\sum_B\pi^*(x,x')} \Big), \notag\\
        & = \ln\Big( (e^\epsilon-1)\frac{\sum_B\pi^*(x,x')}{\sum_A\pi^*(x,x')}+e^\epsilon \Big).\label{eq:initial_parameter}
    \end{align} 
    We get two initial points $\hat{\theta}_a = 1/\Phi$ and $\hat{\theta}_b=n/\Phi$. According to the Brent's method, $[\hat{\theta}_a, \hat{\theta}_b]$ bound the root of $f_{x'}(\theta)$ in each iteration, and in the final iteration, we choose the larger $\hat{\theta}'_b$ as the candidate for given $x'$ to keep $f_{x'}(\hat{\theta}'_b)<0$, which is the relaxed condition of $(\epsilon,\mathbb{S})$-pufferfish privacy in Eq.~\eqref{eq:discrete_relaxed_condition_x_prime}.
\end{proof}
Proposition~\ref{pro:practical_relaxed_mechanism}, modified from Brent's method, provides both theoretical and practical insights for selecting the initial interval of approximate root and determining the parameter $\hat{\theta}$ for all $x'\in\mathcal{X}$. It offers a specific approach to achieving $(\epsilon,\mathbb{S})$-pufferfish privacy for any $\pi^*(x,x')$, any probability distribution, and any given privacy budget $\epsilon$.

\section{Noise Reduction}\label{sec:noise_reduction}
In Proposition~\ref{pro:relaxed_condition_mechanism}, the practical relaxed mechanism alleviates the overly strict condition, and Section~\ref{sec:strict_relaxed_condition} explains the derivation from the strict to relaxed conditions and experimentally visualizes this relaxation. However, a new question arises: 

{\bf Q3.} \textit{Can the practical relaxed mechanism reduce the additive noise compared to the $W_1$ mechanism?}


In this section, we formally introduce the definition and properties of noise reduction.

\subsection{Strict Noise Reduction}
Suppose that, for the $W_1$ mechanism and our proposed practical relaxed mechanism, the additive Laplace noises $N_{\theta_1}$ and $N_{\hat{\theta}}$ provide the same level of $(\epsilon,\mathbb{S})$-pufferfish privacy. Under this equivalence, we define the noise reduction as 
\begin{align*}
    \Delta := \theta_1 - \hat{\theta},    
\end{align*}
which quantifies the reduction in noise scale achieved by the practical relaxed mechanism. 

\begin{theorem}[Strict Noise Reduction]\label{theorem:noise_reduction_proof_general}
    For the noise parameter $\hat{\theta}$ in Proposition~\ref{pro:practical_relaxed_mechanism} and $\theta_1$ in the $W_1$ mechanism~\eqref{eq:w_1_mechanism}, we have
    \begin{enumerate}
        \item {\bf Existence.} There is always a noise reduction $\Delta > 0$ for all privacy budget $\epsilon>0$; \label{theorem:noise_reduction_proof_general_1}
        \item {\bf Trends.} The noise reduction $\Delta$ decreases with $\epsilon$ and becomes significantly large as $\epsilon<1$.\label{theorem:noise_reduction_proof_general_2}
    \end{enumerate}
\end{theorem}
\begin{proof}
For Kantorovich optimal transport plan $\pi^*$ and $n=\max_{(x,x')\in\supp(\pi^*)}|x-x'|$, we have 
\begin{align*}
    f_{x'}(\theta) = \sum_x e^{\frac{|x-x'|}{\theta}} \pi^*(x,x') < e^{\frac{n}{\theta}} \sum_x \pi^*(x,x'),
\end{align*}
where the inequality comes from $e^\frac{1}{\theta} > 1$ and $\pi^*$ are non-negative. 
Then, we define two functions $s_1(z) := \sum_x e^{\frac{|x-x'|}{z}}\pi^*(x,x')$ and $s_2(z) := e^{\frac{n}{z}} \sum_x \pi^*(x,x')$, both are decreasing in $z$. The noise parameter $\hat{\theta}$ and $\theta_1$ are determined by
\begin{align*}
    &\hat{\theta} = s_1^{-1}(e^\epsilon\sum_x\pi^*(x,x')~\text{and}~\theta_1 = s_2^{-1}(e^\epsilon\sum_x\pi^*(x,x')),
\end{align*} 
then we proved $\hat{\theta} < \theta_1 $. This proves Theorem~\ref{theorem:noise_reduction_proof_general}~(\ref{theorem:noise_reduction_proof_general_1}).

To prove Theorem~\ref{theorem:noise_reduction_proof_general}~(\ref{theorem:noise_reduction_proof_general_2}), we first derive the first-order derivative of $\Delta$ and then calculate $\Delta$ when $\epsilon\to0$ as follows.
For $f_{x'}(\hat{\theta})=0$, we take the derivative of $\epsilon$ to get $\Delta'={\theta_1}'-{\hat{\theta}}'$ where ${\theta_1}'=-n/\epsilon^2$ and 
\begin{align*}
    \hat{\theta}' = -\hat{\theta}^2\frac{\sum_x(e^\epsilon\pi^*(x,x'))}{\sum_{x\neq x'}(e^\frac{|x-x'|}{\hat{\theta}}|x-x'|\pi^*(x,x'))}.
\end{align*}
Then, we prove that $\Delta<0$ when $\epsilon<1$. 
Define the support of $\pi^*$ into two space as $A=\{(x,x') \in \supp(\pi^*):|x-x'|\geq 1\}$ and $B=\{(x,x') \in \supp(\pi^*):x=x'\}$, following the initial points in the practical relaxed mechanism in Proposition~\ref{pro:practical_relaxed_mechanism}~(\ref{pro:practical_relaxed_mechanism_1}), we have $\hat{\theta}$ in the interval
\begin{align*}
    \Big[\frac{1}{\underbrace{\ln{\Big((e^\epsilon-1)\frac{\sum_B\pi^*(x,x')}{\sum_A\pi^*(x,x')} + e^\epsilon}\Big)}_{Eq.~\eqref{eq:initial_parameter}}}, ~ \frac{n}{\underbrace{\ln{\Big((e^\epsilon-1)\frac{\sum_B\pi^*(x,x')}{\sum_A\pi^*(x,x')} + e^\epsilon}\Big)}_{Eq.~\eqref{eq:initial_parameter}}}\Big], 
\end{align*}
then the noise reduction $\Delta$ is lower bounded by Eq.~\eqref{eq:mass_pi_lower_bound} and upper bounded by Eq.~\eqref{eq:mass_pi_upper_bound}.
\begin{align}
    \Delta > \frac{n}{\epsilon}-\frac{n}{\ln{\Big((e^\epsilon-1)\frac{\sum_B\pi^*(x,x')}{\sum_A\pi^*(x,x')} + e^\epsilon}\Big)}, \label{eq:mass_pi_lower_bound} \\
    \Delta < \frac{n}{\epsilon}-\frac{1}{\ln{\Big((e^\epsilon-1)\frac{\sum_B\pi^*(x,x')}{\sum_A\pi^*(x,x')} + e^\epsilon}\Big)}. \label{eq:mass_pi_upper_bound}
\end{align}
As $\ln{\Big((e^\epsilon-1)\frac{\sum_B\pi^*(x,x')}{\sum_A\pi^*(x,x')} + e^\epsilon}\Big) > \epsilon$, we have $\lim_{\epsilon\to0}\Delta = \infty$. 
Then we proved that, as the privacy budget $\epsilon$ decreases, the growth rate of $\hat{\theta}$ is slower than that of $\theta_1$. Consequently, the noise reduction, $\Delta = \theta_1-\hat{\theta}$, increases, and $\Delta$ becomes significantly large in the low privacy budget regime.
\end{proof}

Theorem~\ref{theorem:noise_reduction_proof_general}~(\ref{theorem:noise_reduction_proof_general_1}) guarantees that the parameter $\hat{\theta}$ computed by our proposed practical relaxed mechanism (Algorithm~\ref{alg:brent_method}, Proposition~\ref{pro:practical_relaxed_mechanism}) is strictly smaller than $\theta_1$ in the $W_1$ mechanism. Most importantly, this reduction in noise scale is always present, leading to improved data utility. The reason is that it gives a smaller MSE between the released data $Y$ and the original data $X$, as explained in Section~\ref{sec:prelim}. 
The enhanced data utility shows an unnecessary amount of noise by the $W_1$ mechanism and, in return, highlights the advantage of our proposed practical relaxed mechanism. These theoretical results have been visualized in the experimental results in Section~\ref{sec:strict_relaxed_condition}, and the experimental evaluation in Section~\ref{sec:experiment} further verifies the correctness of them, making the findings reliable.

It should be noted that the noise reduction in Theorem~\ref{theorem:noise_reduction_proof_general}~(\ref{theorem:noise_reduction_proof_general_1}) always exists for all privacy budgets. 
In addition, Theorem~\ref{theorem:noise_reduction_proof_general}~(\ref{theorem:noise_reduction_proof_general_2}) implies that the noise reduction $\Delta$ increases as the privacy budget $\epsilon$ becomes smaller, and $\Delta$ is significantly large as $\epsilon<1$, which shows that the data utility gain increases significantly for low privacy budget situations. This meets the demand for data utility while data privacy is tightened in applications.

\paragraph{Example of Theorem~\ref{theorem:noise_reduction_proof_general}}
By Theorem~\ref{theorem:noise_reduction_proof_general}, the strict noise reduction always exists and becomes significant in the low privacy budget regime. To see this clearly, we use a $\pi^*$ such that 
\begin{align*}
    \max_{(x,x') \in \supp(\pi^*)} | x-x'| = 1
\end{align*}
as an example. 
In this case, following Eq.~\eqref{eq:w_1_mechanism} in the $W_1$ mechanism, adding Laplace noise with parameter $\theta_1 = 1/\epsilon$ attains $(\epsilon,\mathbb{S})$-pufferfish privacy.
Following Proposition~\ref{pro:relaxed_condition_mechanism}, solving polynomial equation $f_{x'}(\hat{\theta}) = 0$ for all $x'\in \mathcal{X}$ and determining the maximum yields the parameter:
\begin{align*}
\hat{\theta} = \max_{\rho,(s_i,s_j)\in\mathbb{S}} \sup_{x'\in\mathcal{X}}\frac{1}{\ln\Big(e^{\epsilon} + (e^{\epsilon}-1)\frac{\sum_{B}\pi^*(x,x')}{\sum_{A}\pi^*(x,x')}\Big)},
\end{align*}
where the support of $\pi^*$ is divided into $A := \{(x,x') \in \supp(\pi^*):|x-x'|=1\}$ and $B := \{(x,x') \in \supp(\pi^*) :x=x'\}$. But,
\begin{align}
    \hat{\theta} <  \frac{1}{\ln(e^{\epsilon} )} =  \frac{1}{\epsilon}, \label{eq:noise_reduction_example}
\end{align}
where the inequality follows from the fact that $e^{\epsilon}>1$ and all the $\pi^*$ is non-negative. 
The inequality~\eqref{eq:noise_reduction_example} shows that the noise reduction satisfies 
\begin{align*}
    \Delta = \theta_1 - \hat{\theta} > 0.
\end{align*}
These results further support the correctness of Theorem~\ref{theorem:noise_reduction_proof_general}~(\ref{theorem:noise_reduction_proof_general_1}) and indicate that our proposed mechanism can better preserve data utility than the $W_1$ mechanism~\cite{ding22} when both of them guarantee the same level of pufferfish privacy.

We further construct two prior probability distributions that satisfy $\max_{(x,x') \in \supp(\pi^*)}|x-x'| = 1$ as input (Shown in Table~\ref{tab:dataset_1_1}). 
Following the Eq.~\eqref{eq:w_1_mechanism} in the $W_1$ mechanism and Algorithm~\ref{alg:brent_method} in our proposed practical relaxed mechanism, we calculate the noise parameter $\theta_1$ and $\hat{\theta}$ of the Laplace noise for all privacy budgets $\epsilon\in (0,1]$. The results are shown in Table~\ref{tab:dataset_1_1}. 
%


\begin{table}[ht]
    \caption{Two Prior Probability Distributions such that $\max_{(x,x') \in \supp(\pi^*)}|x-x'| = 1$}
    \label{tab:dataset_1_1}
    \centering
    \begin{tabular}{l l l}
        \toprule
        & $X=0$ & $X=1$ \\
        \midrule
        $P_{X|S}(\cdot|s_i,\rho)$ & 0.52 & 0.48 \\
        $P_{X|S}(\cdot|s_j,\rho)$ & 0.5 & 0.5 \\
        \bottomrule
    \end{tabular}
\end{table}

\begin{figure}[ht]
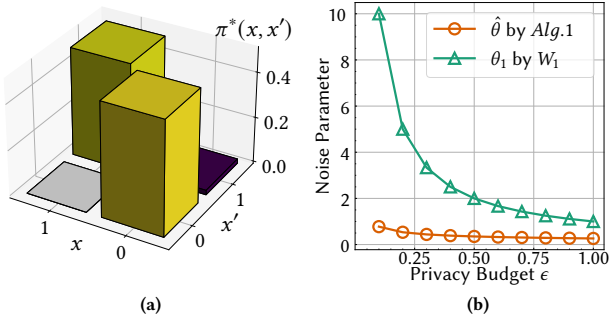

    \centering
    \subfloat[]{%
        \resizebox{0.24\textwidth}{!}{\input{pgf_figs/example_leq_1/optimal_transport_plan_3d.pgf}}
        \label{fig:dataset_example_leq_1_pi}
    }
    \subfloat[]{%
        \resizebox{0.24\textwidth}{!}{\input{pgf_figs/example_leq_1/theta_comparison_new.pgf}}
        \label{fig:dataset_example_leq_1_res}
    }
    \caption{Experimental results using prior distributions in Table~\ref{tab:dataset_1_1}. \ref{fig:dataset_example_leq_1_pi} shows the corresponding Kantorovich optimal transport plan $\pi^*$. \ref{fig:dataset_example_leq_1_res} shows the noise parameter of Laplace noise: $\theta_1$ calibrated by $W_1$ mechanism and $\hat{\theta}$ by our approach in Algorithm~\ref{alg:brent_method}.} 
    \label{fig:dataset_example_leq_1}
\end{figure}

Figure~\ref{fig:dataset_example_leq_1_pi} shows the corresponding Kantorovich optimal transport plan $\pi^*$ where $\max_{(x,x')\in\supp(\pi^*)}|x-x'|=1$. 
And Figure~\ref{fig:dataset_example_leq_1_res} confirms that the noise parameter $\hat{\theta}$ of our proposed practical relaxed mechanism is always smaller than $\theta_1$ of the $W_1$ mechanism for any privacy budget $\epsilon$ (Theorem~\ref{theorem:noise_reduction_proof_general}~(\ref{theorem:noise_reduction_proof_general_1})). 
Notably, for smaller values of $\epsilon$, the noise reduction is particularly significant, which further validates the correctness of Theorem~\ref{theorem:noise_reduction_proof_general}~(\ref{theorem:noise_reduction_proof_general_2}). 
Statistical analysis shows that the noise reduction achieved by the practical relaxed mechanism ranges from 73.5\% to 92.2\% as $\epsilon$ varies within the interval $(0,1]$. 
These results demonstrate that our proposed mechanism consistently reduces the additive noise, thereby improving data utility while ensuring $(\epsilon,\mathbb{S})$-pufferfish privacy. 

\subsection{Variation and Optimality of Noise Reduction}

Below, we analyze how $\Delta$ varies with the probability mass distribution in $\pi^*$ for a given privacy budget $\epsilon$. The purpose is to learn when we can achieve significant noise reduction by replacing the $W_1$ mechanism with our proposed practical relaxed mechanism for attaining $(\epsilon,\mathbb{S})$-pufferfish privacy. 
The lemma below states that the noise reduction $\Delta$ becomes larger if the probability mass is concentrated at the joint elementary events $(x,x')$ that incur lower distances $|x-x'|$. 

\begin{lemma}(Noise Reduction with $\pi^*$)\label{lemma:optimality_noise_reduction}
    For any given privacy budget $\epsilon$, 
    \begin{enumerate}
        \item {\bf Variation.} $\Delta$ is increasing in 
            \begin{align*}
                \frac{\sum_{(x,x')\in\supp(\pi^*), x = x'} \pi^*(x,x')}{\sum_{(x,x')\in\supp(\pi^*), x \neq x'} \pi^*(x,x')}, 
            \end{align*}
        \item {\bf Optimality.} $\Delta$ reaches its maximum $\frac{n}{\epsilon}$ when two prior probability distributions are almost identical like
            \begin{align*}
                \frac{\sum_{(x,x')\in\supp(\pi^*), x = x'} \pi^*(x,x')}{\sum_{(x,x')\in\supp(\pi^*), x \neq x'} \pi^*(x,x')}\to +\infty.
            \end{align*}
    \end{enumerate}    
 
\end{lemma}
\begin{proof}
    Divide the support of $\pi^*$ into two space as $A=\{(x,x') \in \supp(\pi^*):|x-x'|\geq 1\}$ and $B=\{(x,x') \in \supp(\pi^*):x=x'\}$, following the initial points in the practical relaxed mechanism in Proposition~\ref{pro:practical_relaxed_mechanism}~(\ref{pro:practical_relaxed_mechanism_1}), we first derive the exact interval of $\Delta$ and then analyze how this interval evolves as $\frac{\sum_B\pi^*(x,x')}{\sum_A\pi^*(x,x')}$ varies. Then, we analyze the result of $\Delta$ when $\frac{\sum_B\pi^*(x,x')}{\sum_A\pi^*(x,x')}$ approaches infinity.

    The noise reduction $\Delta$ is lower bounded by Eq.~\eqref{eq:mass_pi_lower_bound} and upper bounded by Eq.~\eqref{eq:mass_pi_upper_bound}. 
    When $\pi^*$ assigns more probability mass to data points with low metric distances, $\frac{\sum_B\pi^*(x,x')}{\sum_A\pi^*(x,x')}$ becomes bigger, then $\Delta$ increases.
    When 
    \begin{align*}
        \frac{\sum_{B} \pi^*(x,x')}{\sum_{A} \pi^*(x,x')}\to +\infty,
    \end{align*}
the lower bound \eqref{eq:mass_pi_lower_bound} and upper bound \eqref{eq:mass_pi_upper_bound} of $\Delta$ approach $n/\epsilon$, thus we have $\lim_{\frac{\sum_B\pi^*(x,x')}{\sum_A\pi^*(x,x')}\to+\infty}\Delta = \frac{n}{\epsilon}$.
\end{proof}

Lemma~\ref{lemma:optimality_noise_reduction} shows the situation when our proposed practical relaxed mechanism can significantly reduce the additive Laplace noise while attaining $(\epsilon, \mathbb{S})$-pufferfish privacy.
All the theoretical results in this section demonstrate the improvement in data utility by the noise calibration method in Algorithm~\ref{alg:brent_method} (a.k.a. the practical relaxed mechanism). These results provide a solid theoretical basis for subsequent experimental results in Section~\ref{sec:strict_relaxed_condition} and Section~\ref{sec:experiment}.

\section{\texorpdfstring{Noise Reduction in the Worst Case of $W_1$ Mechanism}{}}\label{sec:w1_equal_dp}

In this section, we analyze the noise reduction when the $W_1$ mechanism is in its worst case, which incurs the largest noise parameter $\theta_1$. We first introduce the worst case of the $W_1$ mechanism and analyze the noise reduction by our proposed practical relaxed mechanism. Moreover, we illustrate that in this case, the $W_1$ mechanism \eqref{eq:w_1_mechanism} can be regarded as an $\ell_1$-sensitivity method (equivalent to~\cite{dmns06} proposed for differential privacy) that measures the maximum pairwise distance in the alphabet $\mathcal{X}^2$. Then, we calculate the noise parameter of these three mechanisms in experiments to show the noise reduction.

\subsection{\texorpdfstring{The worst case of $W_1$ mechanism}{}}\label{sec:worst_case_w_1}
For given $\mathbb{S}$ and $\rho$, consider the $W_1$ mechanism in~\eqref{eq:w_1_mechanism}. There is a chance that 
\begin{equation} \label{eq:W1_worst}
    \max_{(x,x')\in\supp(\pi^*)}|x-x'| = \max_{(x,x')\in\mathcal{X}^2}|x-x'|.
\end{equation}
That is, the Kantorovich optimal transport plan assigns probability mass to a joint elementary event $(x,x')$ that has the maximum pairwise distance in the whole alphabet $\mathcal{X}^2$. 
In this case, $W_1$ mechanism obtains a $\theta_1$ that reaches its highest value. We call this situation the \emph{worst case} of $W_1$ mechanism as it incurs the largest noise for attaining $(\epsilon, \mathbb{S})$-pufferfish privacy. 

Noted that it is not necessary to enforce $\supp(\pi^*) = \mathcal{X}^2$ to satisfy \eqref{eq:W1_worst}.
Proposition~\ref{pro:w1_equal_dp} derives a sufficient condition when the worst-case $W_1$ mechanism will happen in such cases when $\supp(\pi^*) \neq \mathcal{X}^2$ and proposes a specific form of $\pi^*$ in the worst-case $W_1$ mechanism. For simplicity, we assume the alphabet for the prior distributions $P_{X|S}(\cdot|s_i,\rho)$ and $P_{X|S}(\cdot|s_j,\rho)$ is $\mathcal{X}=\{0,1,\dotsc,n\}$.



\begin{proposition}\label{pro:w1_equal_dp}
    For the worst-case $W_1$ mechanism, we have
    \begin{enumerate}
        \item a sufficient condition as
            \begin{align*}
                P_{X|S}(0|s_i,\rho) > 1-P_{X|S}(n|s_j,\rho),
            \end{align*} \label{pro:w1_equal_dp_1}
        \item a specific form of $\pi^*$ as
            \begin{align*}
                \pi^*(x,x') = 
                \begin{cases}
                    P_{X|S}(x'|s_j,\rho), ~ & x = 0 ~\text{and}~  x' \neq n, \\
                    P_{X|S}(x|s_i,\rho), ~ & x \neq 0 ~\text{and}~ x' = n, \\
                    P_{X|S}(0|s_i,\rho) + P_{X|S}(n|s_j,\rho)-1, ~ & x = 0 ~\text{and}~ x' = n, \\
                    0, ~ & \text{others}.
                \end{cases}
            \end{align*} \label{pro:w1_equal_dp_2}
    \end{enumerate}
\end{proposition}
\begin{proof}
    The Kantorovich optimal transport plan $\pi^*$~\cite{vc08, sf15} can be directly determined by
    \begin{align}
        \pi^*(x,x') = \frac{\D^2}{\D x \D x'}\min \Big( F_{s_i}(x), F_{s_j}(x') \Big), \label{eq:calculate_pi_star}
    \end{align}
    where $F_{s_i}(x) = \sum_{k=0}^{x} P_{X|S}(k|s_i,\rho)$ and $F_{s_j}(x') = \sum_{k=0}^{x'} P_{X|S}(k|s_j,\rho)$ are the cumulative mass function (CMF) of the priors $P_{X|S}(x|s_i,\rho)$ and $P_{X|S}(x'|s_j,\rho)$ such that $x,x'\in \{0,1,\dotsc,n\}$.
    To calculate $\pi^*$, we first obtain the joint cumulative mass function of the Kantorovich optimal transport plan 
    \begin{align}
        \pi^*((-\infty,x],(-\infty,x']) = \min\Big( F_{s_i}(x), F_{s_j}(x') \Big) \label{eq:cmf_pi_star}  
    \end{align}
    as follows, 
    \begin{align*}
        & ~ \pi^*((-\infty,0],(-\infty,0]) = \min(P_{X|S}(0|s_i,\rho), P_{X|S}(0|s_j,\rho)) \\
        & ~ \pi^*((-\infty,0],(-\infty,n-1]) = \min(P_{X|S}(0|s_i,\rho), \sum_{x'=0}^{n-1}P_{X|S}(x'|s_j,\rho)) \\
        & ~ \pi^*((-\infty,0],(-\infty,n]) = P_{X|S}(0|s_i,\rho) \\
        & ~ \pi^*((-\infty,n-1],(-\infty,0]) = \min(P_{X|S}(0|s_j,\rho), \sum_{x=0}^{n-1}P_{X|S}(x|s_i,\rho)) \\
        & ~ \pi^*((-\infty,n-1],(-\infty,n]) = \sum_{x=0}^{n-2}P_{X|S}(x|s_i,\rho) \\
        & ~ \pi^*((-\infty,n],(-\infty,0]) = P_{X|S}(0|s_j,\rho)\\
        & ~ \pi^*((-\infty,n],(-\infty,n-1]) = \sum_{x'=0}^{n-1}P_{X|S}(x'|s_j,\rho) \\
        & ~ \pi^*((-\infty,n],(-\infty,n]) = 1
    \end{align*}

    Then, we have the joint probability mass function as\footnote{Recall that for $x_1, x_2, x'_1, x'_2 \in \mathcal{X}$ such that $x_1<x_2$ and $x'_1<x'_2$, $\pi^*([x_1,x_2],[x'_1,x'_2]) = \pi^*((-\infty, x_2],(-\infty,x'_2]) - \pi^*((-\infty, x_1],(-\infty,x'_2]) - \pi^*((-\infty, x_2],(-\infty,x'_1]) + \pi^*((-\infty, x_1],(-\infty,x'_1])$.} 
    \begin{align} 
        \pi^*(x,x') 
        & = \pi^*((-\infty, x]~,(-\infty,x']) - \pi^*((-\infty, x-1]~,(-\infty,x']) \notag\\  
        & - \pi^*((-\infty, x]~,(-\infty,x'-1]) + \pi^*((-\infty, x]~,(-\infty,x']). \label{eq:pmf_pi_star}
    \end{align}
    
    If the $W_1$ mechanism is in the worst case, we have
    \begin{align*}
        \max_{x,x'\in \supp{(\pi^*)}}|x-x'| = n \Rightarrow
        \pi^*(0,n) \neq 0 ~\text{or}~ \pi^*(n,0) \neq 0.
    \end{align*}
    Due to the symmetry of $x$ and $x'$, we only consider $\pi^*(0,n) \neq 0$ and then we have
    \begin{align*}
        & P_{X|S}(0|s_i,\rho) - \min\Big(P_{X|S}(0|s_i,\rho), \sum_{x'=0}^{n-1}P_{X|S}(x'|s_j,\rho)\Big)\neq 0.
    \end{align*}
    Thus, we prove Proposition~\ref{pro:w1_equal_dp} (\ref{pro:w1_equal_dp_1}): if the following condition holds
    \begin{align}
        P_{X|S}(0|s_i,\rho) > 1-P_{X|S}(n|s_j,\rho) \label{eq:sufficient_condition_worst_case_w1}
    \end{align}
    The $W_1$ mechanism reaches its worst case.

    Following Eq.~\eqref{eq:sufficient_condition_worst_case_w1}, we derive $\pi^*((-\infty,x],(-\infty,x'])$, which is shown in Matrix~\eqref{eq:matrix_pi_infinite}.
    \begin{align}\label{eq:matrix_pi_infinite}
        \begin{bmatrix}
        P_{X|S}(0|s_j,\rho) & \cdots & \sum_{x'=0}^{n-1} P_{X|S}(x'|s_j,\rho) & P_{X|S}(0|s_i,\rho) \\
        \vdots & \ddots & \vdots & \vdots \\
        P_{X|S}(0|s_j,\rho) & \cdots & \sum_{x'=0}^{n-1} P_{X|S}(x'|s_j,\rho) & \sum_{x=0}^{k} P_{X|S}(x|s_i,\rho) \\
        \vdots & \ddots & \vdots & \vdots \\
        P_{X|S}(0|s_j,\rho) & \cdots & \sum_{x'=0}^{n-1} P_{X|S}(x'|s_j,\rho) & \sum_{x=0}^{n-1} P_{X|S}(x|s_i,\rho) \\
        P_{X|S}(0|s_j,\rho) & \cdots & \sum_{x'=0}^{n-1} P_{X|S}(x'|s_j,\rho) & 1
        \end{bmatrix}    
    \end{align}
    Then, we have $\pi^*(x,x')$ in Matrix~\eqref{eq:matrix_pi_star}, like
    \begin{align}\label{eq:matrix_pi_star}
        \begin{bmatrix}
        P_{X|S}(0|s_j,\rho)   & \cdots & P_{X|S}(n-1|s_j,\rho)   & \substack{P_{X|S}(0|s_i,\rho) \\ + P_{X|S}(n|s_j,\rho) - 1} \\
        0                & \cdots & 0                  & P_{X|S}(1|s_i,\rho) \\
        \vdots           & \ddots & \vdots             & \vdots \\
        0                & \cdots & 0                  & P_{X|S}(n-1|s_i,\rho) \\
        0                & \cdots & 0                  & P_{X|S}(n|s_i,\rho)
        \end{bmatrix}.
    \end{align}
    Thus, we can formulate it as
    \begin{align*}
        \pi^*(x,x') = 
        \begin{cases}
            P_{X|S}(x'|s_j,\rho) & x = 0 ~\mathrm{and}~  x' \neq n, \\
            P_{X|S}(x|s_i,\rho) & x \neq 0 ~\mathrm{and}~ x' = n, \\
            P_{X|S}(0|s_i,\rho)+P_{X|S}(n|s_j,\rho)-1 & x = 0 ~\mathrm{ and }~ x' = n, \\
            0 & \mathrm{others}.
        \end{cases}
    \end{align*}
    Then Proposition~\ref{pro:w1_equal_dp}~(\ref{pro:w1_equal_dp_2}) is proved.
\end{proof}

Following the specific form in Proposition~\ref{pro:w1_equal_dp}~(\ref{pro:w1_equal_dp_2}), we get the $\pi^*$ and then derive the noise parameter $\theta_1$ for the $W_1$ mechanism~\eqref{eq:w_1_mechanism}. Thus, we propose the empirical conditions, following which our practical relaxed mechanism achieves a large noise reduction compared to the worst-case $W_1$ mechanism.
\begin{remark}\label{remark:noise_reduction_kan_equal_dp}
    The noise reduction achieved by the practical relaxed mechanism is significant if all the following conditions are satisfied.
    \begin{enumerate}
        \item $\delta := P_{X|S}(0|s_i,\rho) - (1-P_{X|S}(n|s_j,\rho)) > 0$,  \label{remark:noise_reduction_kan_equal_dp_1}
        \item $\delta\to 0^+$ and $P_{X|S}(x|s_i,\rho)<\delta, ~P_{X|S}(x'|s_j,\rho)<\delta, ~\forall x,x'\in\{1,...,n-1\}$, \label{remark:noise_reduction_kan_equal_dp_2}
    \end{enumerate}
\end{remark}
Remark~\ref{remark:noise_reduction_kan_equal_dp}~(\ref{remark:noise_reduction_kan_equal_dp_1}) recalls the results in Proposition~\ref{pro:w1_equal_dp} to ensure the worst case of the $W_1$ mechanism, where Eq.~\eqref{eq:W1_worst} holds.
And Remark~\ref{remark:noise_reduction_kan_equal_dp}~(\ref{remark:noise_reduction_kan_equal_dp_2}) recalls the results we proposed in Lemma~\ref{lemma:optimality_noise_reduction}, which makes 
\begin{align*}
    \frac{\sum_{(x,x')\in\supp(\pi^*), x = x'} \pi^*(x,x')}{\sum_{(x,x')\in\supp(\pi^*), x \neq x'} \pi^*(x,x')} = \frac{P_{X|S}(0|s_j,\rho) + P_{X|S}(n|s_i,\rho)}{1-P_{X|S}(0|s_j,\rho) - P_{X|S}(n|s_i,\rho)}
\end{align*}
bigger to achieve the larger noise reduction. 
All these remarks show that a large noise reduction can be achieved by our proposed practical relaxed mechanism compared to the worst-case $W_1$ mechanism. We will construct two prior probability distributions satisfying Remark~\ref{remark:noise_reduction_kan_equal_dp} and conduct experiments to verify this in Section~\ref{sec:simulation_worst_case}.

\paragraph{Equivalence to $\ell_1$-sensitivity Method} Recall the $\ell_1$-sensitivity method that was originally proposed in~\cite{dmns06} for attaining $\epsilon$-differential privacy. We describe the corresponding scenario in pufferfish privacy setting as follows. 
Let $D|s_i$ and $D|s_j$ be the two different databases resulting from secrets $s_i$ and $s_j$, respectively. Here, $s_i$ and $s_j$ could refer to an individual's existence and nonexistence, respectively, in the system. However, there is no restriction that $D|s_i$ and $D|s_j$ differ in only one entry, i.e., pufferfish privacy~\cite{da14} disregards the database neighborhood constraint. Assume the adversary repeatedly queries the database to obtain the aggregated statistics. 

For query function $q(\cdot)$, let $X = q(D)$ be the query answers of database $D$. Unlike Differential Privacy, we assume that $X$ is a random variable that depends on the secret $S$, where the randomness could arise from $q$ (e.g., a randomized query function) or $D$ (i.e., there are some probability distributions governing the appearance frequency for each database). This results in priors $P_{X|S}(\cdot|s_i,\rho)$ and $P_{X|S}(\cdot|s_j,\rho)$ given secret instance $s_i$ and $s_j$, respectively. 
Assume the worst case condition~\eqref{eq:W1_worst} satisfies, i.e., there exists $(x,x') \in \supp(\pi^*)$ such that $|x-x'| = \max_{(x,x')\in \mathcal{X}^2} |x-x'| = n$. This necessarily means that $P_{X|S}(x|s_i,\rho) > 0$ and $P_{X|S}(x'|s_i,\rho) > 0$. That is, query answers $q(D|s_i) = x$ and $q(D|s_j) = x'$ appear at least once if the adversary queries a sufficient number of times.

Instead of $W_1$ mechanism, consider the $\ell_1$-sensitivity method, which disregards the probability distribution of the query answer $X$, but straightforwardly calculates the maximum distance in $X$ between $s_i$ and $s_j$: $\max_{(x,x') \in \mathcal{X}^2}|X|s_i - X'|s_j| = \max |q(D|s_i) - q(D|s_j)|$. This is called $\ell_1$-sensitivity of the query function $q$ and the $\ell_1$-sensitivity equals $n$ in the worst case $W_1$ mechanism. The noise parameters resulting from both methods are the same: 
\begin{align}
    \theta_{\ell} = \frac{ \max_{(x,x') \in \mathcal{X}^2}|X|s_i - X'|s_j| }{\epsilon} = \frac{n}{\epsilon}, \label{eq:theta_ell_1} 
    \\ = \frac{\max_{(x,x') \in \supp(\pi^*)} |x-x'|}{\epsilon} = \theta_1 .\notag
\end{align}

In this case, the $W_1$ mechanism that takes into account the prior distributions reduces to $\ell_1$ method regardless of the intrinsic randomness in $X$. 
Knowing that differential privacy is a special case of pufferfish privacy~\cite{da14} where it usually consumes the largest noise amount to attain data security~\cite{ding22}, the above equivalence to the $\ell_1$-sensitivity method (for differential privacy) in return explains why the $W_1$ mechanism satisfying \eqref{eq:W1_worst} is so called the worst case.

\subsection{Experiments on Constructed Distributions}\label{sec:simulation_worst_case}
\begin{figure}[ht]
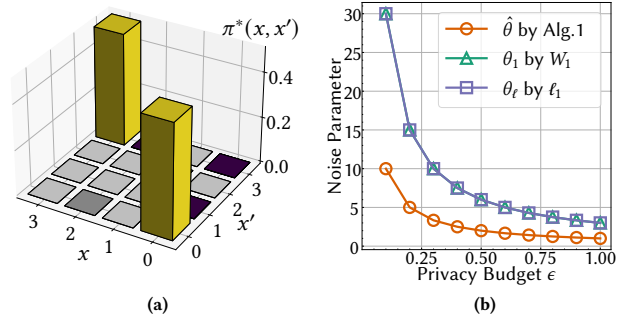

    \centering
    \subfloat[]{%
        \resizebox{0.24\textwidth}{!}{\input{pgf_figs/w1_equal_dp/optimal_transport_plan_3d.pgf}}
        \label{fig:dataset_w1_equal_dp_pi}
    }
    \subfloat[]{%
        \resizebox{0.24\textwidth}{!}{\input{pgf_figs/w1_equal_dp/theta_comparison_new.pgf}}
        \label{fig:dataset_w1_equal_dp_res}
    }
    \caption{Experimental results in the worst case $W_1$ mechanism. \ref{fig:dataset_w1_equal_dp_pi} shows the corresponding Kantorovich optimal transport plan $\pi^*$ where $\pi^*(0,3) \neq 0$ satisfying condition \eqref{eq:W1_worst}. \ref{fig:dataset_w1_equal_dp_res} shows the noise parameter of Laplace noise: $\theta_1$ calibrated by $W_1$ mechanism, $\theta_{\ell}$ by $\ell_1$-sensitivity method, and $\hat{\theta}$ by our proposed practical relaxed mechanism in Algorithm~\ref{alg:brent_method}.}
    \label{fig:dataset_w1_equal_dp}
\end{figure}

We construct two prior probability distributions in Table~\ref{tab:dataset_4_5} giving rise to the worst case $W_1$ mechanism, as stated in Proposition~\ref{pro:w1_equal_dp} and Remark~\ref{remark:noise_reduction_kan_equal_dp},  and calculate the noise parameters $\theta_1$ (Eq.~\eqref{eq:w_1_mechanism}) for $W_1$ mechanism, $\theta_\ell$ (Eq.~\eqref{eq:theta_ell_1}) for $\ell_1$-sensitivity method, and $\hat{\theta}$ (Algorithm~\ref{alg:brent_method}) for our proposed practical relaxed mechanism. The results are shown in Figure~\ref{fig:dataset_w1_equal_dp} and Table~\ref{tab:dataset_results}.

\begin{table}[!ht]
    \caption{Two Prior Probability Distributions in the Worst-Case $W_1$ Mechanism Satisfying Remark~\ref{remark:noise_reduction_kan_equal_dp}}
    \label{tab:dataset_4_5}
    \centering
    \begin{tabular}{l l l l l}
        \toprule
        & $X=0$ & $X=1$ & $X=2$ & $X=3$ \\
        \midrule
        $P_{X|S}(\cdot|s_i,\rho)$ & 0.50001 & 0 & 0.00001 & 0.49998 \\
        $P_{X|S}(\cdot|s_j,\rho)$ & 0.49996 & 0.00001 & 0 & 0.50003 \\
        \bottomrule
    \end{tabular}
\end{table}

As can be seen from Figure~\ref{fig:dataset_w1_equal_dp_res}, $\theta_1$ for the $W_1$ mechanism always equals to $\theta_{\ell}$ for the $\ell_1$-sensitivity method ($\theta_1$ overlaps $\theta_{\ell}$ in the figure). 
By comparison, $\hat{\theta}$ of our approach is smaller for all privacy budgets (Theorem~\ref{theorem:noise_reduction_proof_general}~({\ref{theorem:noise_reduction_proof_general_1}})), achieving the noise reduction of approximately $66.7\%$ when $\epsilon$ lies in the range $(0,1]$. And the reduction in noise becomes significant large in a low privacy budget (Theorem~\ref{theorem:noise_reduction_proof_general}~({\ref{theorem:noise_reduction_proof_general_2}})). They validate Remark~\ref{remark:noise_reduction_kan_equal_dp} and our following analysis, and highlight the advantage of our proposed mechanism in preserving data utility for attaining $(\epsilon,\mathbb{S})$-pufferfish privacy.

\section{Real-World Experiments}\label{sec:experiment}
We run experiments on three 
real-world datasets in the UCI machine learning repository~\cite{mrk} to validate the results and findings we derived in this paper.

\begin{table*}[ht]
    \caption{Dataset Attributes Informations}
    \label{tab:dataset_experiment}
    \centering
    \begin{tabular}{l c c c c}
        \toprule
        Dataset & Instances & Sensitive attribute & Public attribute & Support of $X$ \\
        \midrule
        \textit{Student Performance}~\cite{student_performance} & 649 & `higher' & `romantic' & $\mathcal{X}=\{0,1\}$ \\
        \textit{Census Income}~\cite{census_income} & 48842 & `marital-status' & `workclass' & $\mathcal{X}=\{0,1,\cdots,7,8\}$ \\
        \textit{Bank Marketing}~\cite{bank_marketing} & 45211 & `loan' & `marital' & $\mathcal{X}=\{0,1,2\}$ \\
        \bottomrule
    \end{tabular}
\end{table*}

\begin{table*}[!ht]
    \caption{Detailed Results}
    \label{tab:dataset_results}
    \centering
    \begin{tabular}{l | l | c c c c c c c c c c}
        \toprule
        Dataset & Mechanism & $\epsilon=0.1$ & $\epsilon=0.2$ & $\epsilon=0.3$ & $\epsilon=0.4$ & $\epsilon=0.5$ & $\epsilon=0.6$ & $\epsilon=0.7$ & $\epsilon=0.8$ & $\epsilon=0.9$ & $\epsilon=1.0$ \\
        \midrule
        \multirow{3}{*}{\textit{Simulation (Figure~\ref{fig:dataset_example_leq_1_res})}} & 
        $\ell_1$~in~Eq.\eqref{eq:theta_ell_1} & 10.00 & 5.00 & 3.33 & 2.50 & 2.00 & 1.67 & 1.43 & 1.25 & 1.11 & 1.00  \\ 
        & $W_1$~\cite{ding22} & 10.00 & 5.00 & 3.33 & 2.50 & 2.00 & 1.67 & 1.43 & 1.25 & 1.11 & 1.00  \\ 
        & \cellcolor[gray]{0.9} {\bf Alg.~\ref{alg:brent_method}} & \cellcolor[gray]{0.9} {\bf 0.78} & \cellcolor[gray]{0.9} {\bf 0.54} & \cellcolor[gray]{0.9} {\bf 0.44} & \cellcolor[gray]{0.9} {\bf 0.39} & \cellcolor[gray]{0.9} {\bf 0.35} & \cellcolor[gray]{0.9} {\bf 0.33} & \cellcolor[gray]{0.9} {\bf 0.31} & \cellcolor[gray]{0.9} {\bf 0.29} & \cellcolor[gray]{0.9} {\bf 0.28} & \cellcolor[gray]{0.9} {\bf 0.26}  \\ 
        \hline
        \multirow{3}{*}{\textit{Simulation (Figure~\ref{fig:dataset_w1_equal_dp_res})}} & 
        $\ell_1$~in~Eq.\eqref{eq:theta_ell_1} & 30.00 & 15.00 & 10.00 & 7.50 & 6.00 & 5.00 & 4.29 & 3.75 & 3.33 & 3.00  \\ 
        & $W_1$~\cite{ding22} & 30.00 & 15.00 & 10.00 & 7.50 & 6.00 & 5.00 & 4.29 & 3.75 & 3.33 & 3.00  \\ 
        & \cellcolor[gray]{0.9} {\bf Alg.~\ref{alg:brent_method}} & \cellcolor[gray]{0.9} {\bf 10.00} & \cellcolor[gray]{0.9} {\bf 5.00} & \cellcolor[gray]{0.9} {\bf 3.33} & \cellcolor[gray]{0.9} {\bf 2.50} & \cellcolor[gray]{0.9} {\bf 2.00} & \cellcolor[gray]{0.9} {\bf 1.67} & \cellcolor[gray]{0.9} {\bf 1.43} & \cellcolor[gray]{0.9} {\bf 1.25} & \cellcolor[gray]{0.9} {\bf 1.11} & \cellcolor[gray]{0.9} {\bf 1.00}  \\ 
        \hline
        \multirow{3}{*}{\textit{Student Performance}~\cite{student_performance}} & $\ell_1$~in~Eq.\eqref{eq:theta_ell_1} & 10.00 & 5.00 & 3.33 & 2.50 & 2.00 & 1.67 & 1.43 & 1.25 & 1.11 & 1.00  \\ 
        & $W_1$~\cite{ding22} & 10.00 & 5.00 & 3.33 & 2.50 & 2.00 & 1.67 & 1.43 & 1.25 & 1.11 & 1.00  \\ 
        & \cellcolor[gray]{0.9} {\bf Alg.~\ref{alg:brent_method}} & \cellcolor[gray]{0.9} {\bf 3.39} & \cellcolor[gray]{0.9} {\bf 1.84} & \cellcolor[gray]{0.9} {\bf 1.31} & \cellcolor[gray]{0.9} {\bf 1.04} & \cellcolor[gray]{0.9} {\bf 0.88} & \cellcolor[gray]{0.9} {\bf 0.77} & \cellcolor[gray]{0.9} {\bf 0.68} & \cellcolor[gray]{0.9} {\bf 0.62} & \cellcolor[gray]{0.9} {\bf 0.57} & \cellcolor[gray]{0.9} {\bf 0.53}  \\ 
        \hline
        \multirow{3}{*}{\textit{Census Income}~\cite{census_income}} & 
        $\ell_1$~in~Eq.\eqref{eq:theta_ell_1} & 80.00 & 40.00 & 26.67 & 20.00 & 16.00 & 13.33 & 11.43 & 10.00 & 8.89 & 8.00 \\ 
        & $W_1$~\cite{ding22} & 20.00 & 10.00 & 6.67 & 5.00 & 4.00 & 3.33 & 2.86 & 2.50 & 2.22 & 2.00  \\ 
        & \cellcolor[gray]{0.9} {\bf Alg.~\ref{alg:brent_method}} & \cellcolor[gray]{0.9} {\bf 10.00} & \cellcolor[gray]{0.9} {\bf 5.00} & \cellcolor[gray]{0.9} {\bf 3.33} & \cellcolor[gray]{0.9} {\bf 2.50} & \cellcolor[gray]{0.9} {\bf 2.05} & \cellcolor[gray]{0.9} {\bf 1.76} & \cellcolor[gray]{0.9} {\bf 1.54} & \cellcolor[gray]{0.9} {\bf 1.38} & \cellcolor[gray]{0.9} {\bf 1.25} & \cellcolor[gray]{0.9} {\bf 1.15}  \\ 
        \hline
        \multirow{3}{*}{\textit{Bank Marketing}~\cite{bank_marketing}} & 
        $\ell_1$~in~Eq.\eqref{eq:theta_ell_1} & 20.00 & 10.00 & 6.67 & 5.00 & 4.00 & 3.33 & 2.86 & 2.50 & 2.22 & 2.00  \\ 
        & $W_1$~\cite{ding22} & 10.00 & 5.00 & 3.33 & 2.50 & 2.00 & 1.67 & 1.43 & 1.25 & 1.11 & 1.00  \\ 
        & \cellcolor[gray]{0.9} {\bf Alg.~\ref{alg:brent_method}} & \cellcolor[gray]{0.9} {\bf 2.53} & \cellcolor[gray]{0.9} {\bf 1.42} & \cellcolor[gray]{0.9} {\bf 1.04} & \cellcolor[gray]{0.9} {\bf 0.84} & \cellcolor[gray]{0.9} {\bf 0.72} & \cellcolor[gray]{0.9} {\bf 0.64} & \cellcolor[gray]{0.9} {\bf 0.58} & \cellcolor[gray]{0.9} {\bf 0.53} & \cellcolor[gray]{0.9} {\bf 0.49} & \cellcolor[gray]{0.9} {\bf 0.46}  \\ 
        \bottomrule
    \end{tabular}
\end{table*}

\begin{itemize}
    \item \textit{Student Performance}~\cite{student_performance}. The dataset contains student achievement data in secondary education collected from Portuguese schools.
    \item \textit{Census Income}~\cite{census_income}. The dataset predicts whether the annual income of an individual exceeds \$50K/yr.
    \item \textit{Bank Marketing}~\cite{bank_marketing}. The dataset is related to direct marketing campaigns of a Portuguese banking institution.
\end{itemize}

\subsection{Experiment Settings}

\begin{figure*}[t]
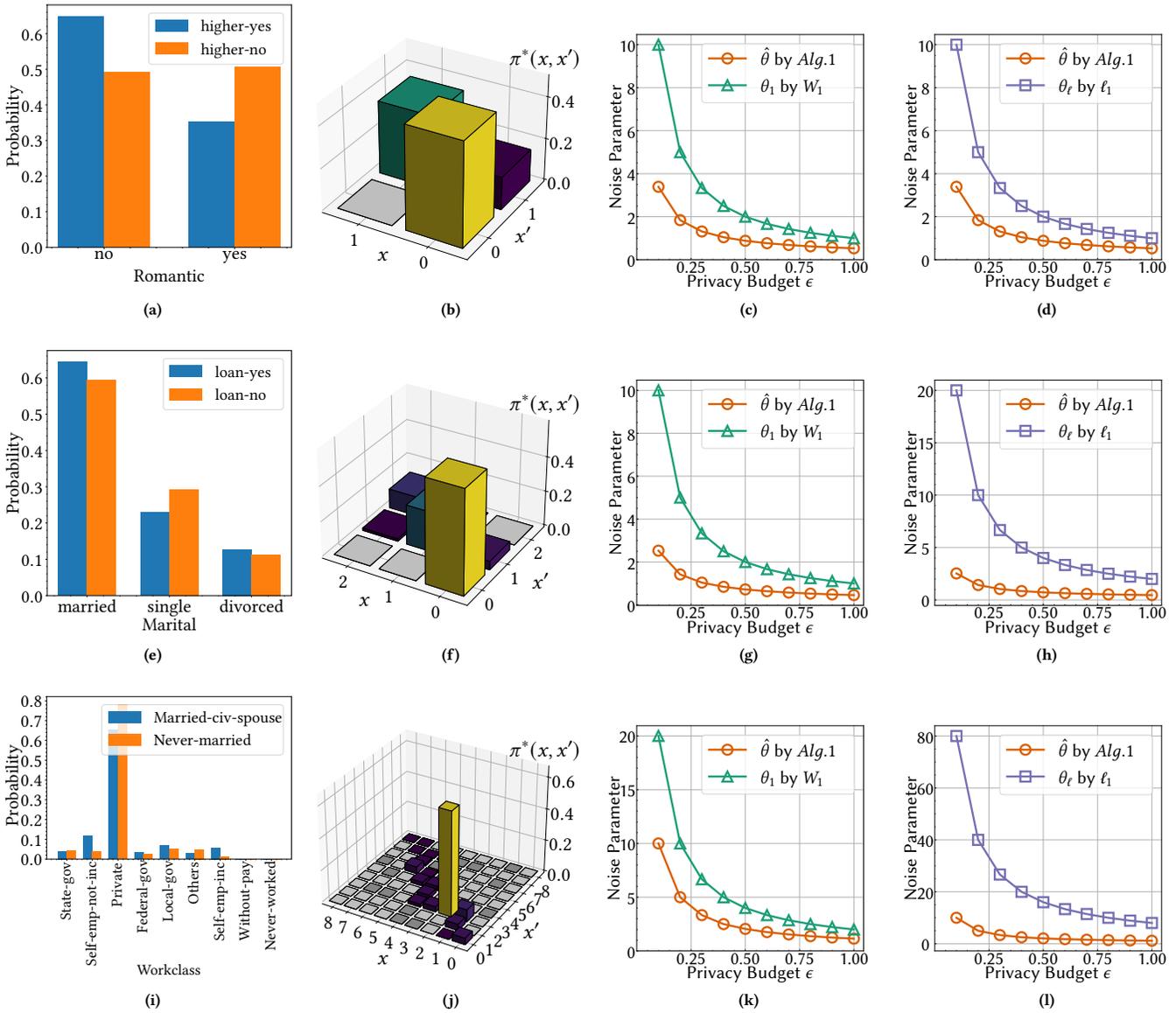

    \centering
    \subfloat[]{%
        \resizebox{0.25\textwidth}{!}{\input{pgf_figs/student/probability_distribution_student.pgf}}
        \label{fig:dataset_student_probability}
    }
    \subfloat[]{%
        \resizebox{0.25\textwidth}{!}{\input{pgf_figs/student/optimal_transport_plan_3d.pgf}}
        \label{fig:dataset_student_pi}
    }
    \subfloat[]{%
        \resizebox{0.25\textwidth}{!}{\input{pgf_figs/student/theta_comparison_new.pgf}}
        \label{fig:dataset_student_res}
    }
    \subfloat[]{%
        \resizebox{0.25\textwidth}{!}{\input{pgf_figs/student/theta_comparison_ell.pgf}}
        \label{fig:dataset_student_res_ell}
    }
    \\
    \subfloat[]{%
        \resizebox{0.25\textwidth}{!}{\input{pgf_figs/bank/probability_distribution_bank.pgf}}
        \label{fig:dataset_bank_probability}
    }
    \subfloat[]{%
        \resizebox{0.25\textwidth}{!}{\input{pgf_figs/bank/optimal_transport_plan_3d.pgf}}
        \label{fig:dataset_bank_pi}
    }
    \subfloat[]{%
        \resizebox{0.25\textwidth}{!}{\input{pgf_figs/bank/theta_comparison_new.pgf}}
        \label{fig:dataset_bank_res}
    }
    \subfloat[]{%
        \resizebox{0.25\textwidth}{!}{\input{pgf_figs/bank/theta_comparison_ell.pgf}}
        \label{fig:dataset_bank_res_ell}
    }
    \\
    \subfloat[]{%
        \resizebox{0.25\textwidth}{!}{\input{pgf_figs/census/probability_distribution_census.pgf}}
        \label{fig:dataset_census_probability}
    }
    \subfloat[]{%
        \resizebox{0.25\textwidth}{!}{\input{pgf_figs/census/optimal_transport_plan_3d.pgf}}
        \label{fig:dataset_census_pi}
    }
    \subfloat[]{%
        \resizebox{0.25\textwidth}{!}{\input{pgf_figs/census/theta_comparison_new.pgf}}
        \label{fig:dataset_census_res}
    }
    \subfloat[]{%
        \resizebox{0.25\textwidth}{!}{\input{pgf_figs/census/theta_comparison_ell.pgf}}
        \label{fig:dataset_census_res_ell}
    }
    \caption{
    In \textit{Student Performance}: \ref{fig:dataset_student_probability} shows the prior distributions of ‘Romantic’ conditioned on ‘higher-yes’ and `higher-no' events. \ref{fig:dataset_student_pi} illustrates the corresponding Kantorovich optimal transport plan $\pi^*$. \ref{fig:dataset_student_res} and \ref{fig:dataset_student_res_ell} show the Laplace noise parameter, $\theta_1$ by $W_1$ mechanism, $\theta_\ell$ by $\ell_1$-sensitivity mechanism and $\hat{\theta}$ by Algorithm~\ref{alg:brent_method}. \\
    In \textit{Bank Marketing}: \ref{fig:dataset_student_probability} shows the prior distributions of ‘Marital’ conditioned on ‘loan-yes’ and `loan-no' events. \ref{fig:dataset_bank_pi} shows the corresponding Kantorovich optimal transport plan $\pi^*$. \ref{fig:dataset_bank_res} and \ref{fig:dataset_bank_res_ell} present the Laplace noise parameter, $\theta_1$ by $W_1$ mechanism, $\theta_\ell$ by $\ell_1$-sensitivity mechanism and $\hat{\theta}$ by Algorithm~\ref{alg:brent_method}. \\
    In \textit{Census Income}: \ref{fig:dataset_student_probability} shows the prior distributions of ‘Workclass’ conditioned on ‘Married-civ-spouse’ and `Never-married' events. \ref{fig:dataset_census_pi} illustrates the corresponding Kantorovich optimal transport plan $\pi^*$. \ref{fig:dataset_census_res} and \ref{fig:dataset_census_res_ell} show the Laplace noise parameter, $\theta_1$ by $W_1$ mechanism, $\theta_\ell$ by $\ell_1$-sensitivity mechanism and $\hat{\theta}$ by Algorithm~\ref{alg:brent_method}.
    }
\end{figure*}

\paragraph{Attributes} 
In our experiments, we assume scenarios in which adversaries can infer sensitive attributes by querying public attributes, leveraging prior knowledge of the adversaries.
In \textit{Student Performance}, we focus on two attributes: `higher' and `romantic'. $S$ refers to the sensitive attribute `higher', which represents whether a student wants to take higher education, and $X$ refers to the attribute `romantic', which represents whether a student has a romantic relationship. The correlation is that the students who don't have a romantic relationship have more passion to take higher education. 
We want to publish the column `romantic' while protecting the privacy of `higher' for all students. The probability distributions are shown in Figure~\ref{fig:dataset_student_probability}.  
We redefine the `higher' degree from $X=0$ to $X=1$, which represents `yes' or `no' in the `higher' attribute. Consider the events `higher-yes' denoted as $s_i$ and `higher-no' denoted as $s_j$. The support of $P_{X|S}(\cdot|s_i, \rho)$ and $P_{X|S}(\cdot|s_j, \rho)$ is $\mathcal{X}=\{0,1\}$. 
What's more, the sensitive and public attributes are `marital-status' and `workclass' in the \textit{Census Income} dataset (As shown in Figure~\ref{fig:dataset_census_probability}). 
In the \textit{Bank Marketing} dataset, `loan' and `marital' are the sensitive and public columns (As shown in Figure~\ref{fig:dataset_bank_probability}). 
Overall, all the attributes and prior probability distributions we used in the experiments are shown in Table~\ref{tab:dataset_experiment}.

\paragraph{Methodologies} 
In each real-world dataset, we first compute the prior probability distributions $P_{X|S}(\cdot|s_i, \rho)$ and $P_{X|S}(\cdot|s_j, \rho)$ for $\mathcal{X} = \{0,1,\dotsc,n\}$
under two sensitive attributes $s_i$ and $s_j$. 
The Kantorovich optimal transport plan $\pi^*$ can be directly determined by Eq.~\eqref{eq:calculate_pi_star}. In detail, we first obtain the joint cumulative mass function $\pi^*((-\infty,x],(-\infty,x'])$ (Eq.~\eqref{eq:cmf_pi_star}) and then join probability mass function $\pi^*(x,x')$ (Eq.~\eqref{eq:pmf_pi_star}). 
Finally, we compare our proposed practical relaxed mechanism with two previous noise calibration methods: the $\ell_1$-sensitivity method~\eqref{eq:theta_ell_1} and the $W_1$ mechanism~\eqref{eq:w_1_mechanism}. The noise parameter is obtained over a privacy budget range $\epsilon$ in (0, 1], which represents a high privacy guarantee in applications.

\paragraph{Evaluation} 
The additive noise in all mechanisms we used in this paper follows a Laplace distribution, with variance $2\theta^2$, which is also the mean squared error (MSE) between the released and the original data (see Eq.~\eqref{eq:VAR})
Thus, a smaller $\theta$ implies smaller MSE between released and original data, also higher utility. In our experiments, we use the noise parameter $\theta$ as an indicator to evaluate the performance in data utility for different mechanisms.

\subsection{Results}
We analyze the experimental results in these three real-world datasets and verify the theoretical results about the advantages in the noise reduction of our proposed practical relaxed mechanism and the trends of noise reduction with different privacy budget $\epsilon$.

\paragraph{Existence of Noise Reduction}
To show the advantages of our proposed practical relaxed mechanism in the noise reduction over other mechanisms while maintaining the same level of pufferfish privacy, we conduct experiments on three real-world datasets. We analyze the results on each dataset separately as follows. 

In the \textit{Student Performance} dataset, the two prior probability distributions $P_{X|S}(\cdot|s_i,\rho)$ and $P_{X|S}(\cdot|s_j,\rho)$, which are shown in Figure~\ref{fig:dataset_student_probability}, represent the probabilities of a student being in a romantic relationship, conditioned on two sensitive attributes $S$--specifically, whether the student wants to pursue higher education or not. 
We calculate the Kantorovich optimal transport plan $\pi^*$ and plot the $\pi^*(x,x')$ in Figure~\ref{fig:dataset_student_pi}. 
For each mechanism, we evaluate across the range of privacy budgets $\epsilon$ in $(0, 1]$. 
As illustrated in Figure~\ref{fig:dataset_student_res}, Figure~\ref{fig:dataset_student_res_ell} and Table~\ref{tab:dataset_results}, our proposed mechanism consistently yields a smaller noise parameter $\hat{\theta}$ than the $W_1$ mechanism and $\ell_1$-sensitivity method. 
In particular, it reduces the noise parameter $\hat{\theta}$ by approximately \textbf{66\%} to \textbf{47\%} compared to the $W_1$ mechanism and the $\ell_1$-sensitivity method.  
The experimental results show that the practical relaxed mechanism reduces noise while preserving privacy, therefore maintaining data utility. 

Then, we replicate the same experimental procedure on the \textit{Bank Marketing} and \textit{Census Income} datasets. 
In the \textit{Bank Marketing} dataset, our proposed mechanism achieves a reduction in noise parameter from \textbf{75\%} to \textbf{54\%} from the $W_1$ mechanism and achieves \textbf{87\%} to \textbf{77\%} reduction from the $\ell_1$-sensitivity method, which are consistent with our findings above. 
In the \textit{Census Income} dataset, our approach demonstrates improvements from \textbf{50\%} to \textbf{43\%} in the noise parameter than the $W_1$ mechanism and from \textbf{88\%} to \textbf{86\%} from the $\ell_1$-sensitivity method as the privacy budget $\epsilon$ decreases, further highlighting the better performance of our mechanism in balancing privacy and utility.

\paragraph{Trends of Noise Reduction}
In these three real-world datasets, the noise parameters $\theta$ and $\hat{\theta}$ increase with the privacy budget $\epsilon$ as it becomes small. Furthermore, the noise reduction $\theta-\hat{\theta}$, denoted as $\Delta$, becomes larger with the $\epsilon$ as it becomes small. These experimental results verify the correctness of our proposed theoretical results, highlighting the advantages of our proposed practical relaxed mechanism to attain pufferfish privacy.

\paragraph{\texorpdfstring{Noise Reduction in the Worst-Case $W_1$ Mechanism}{}}
In Section~\ref{sec:w1_equal_dp}, we analyze the noise reduction in the worst case of the $W_1$ mechanism, where the noise parameter of the $W_1$ mechanism is equal to that of the $\ell_1$-sensitivity method~\cite{dmns06}. 
In the experimental results (Figure~\ref{fig:dataset_student_res}, Figure~\ref{fig:dataset_student_res_ell} and Table~\ref{tab:dataset_results}), the prior probability distribution and its $\pi^*$ in \textit{Student Performance} satisfy the worst case of the $W_1$ mechanism. The experimental results show that the noise parameter of the $W_1$ mechanism is always equal to that of the $\ell_1$-sensitivity method in all privacy budgets $\epsilon$, and our proposed practical relaxed mechanism still achieves the noise reduction by approximately \textbf{66\%} to \textbf{47\%}. They verify our results in Section~\ref{sec:w1_equal_dp}.

\section{Conclusion}\label{sec:conclusion}
In this paper, we proposed a practical relaxed mechanism to enhance data utility for attaining pufferfish privacy, which is based on the existing Wasserstein/Kantorovich mechanism by alleviating its overly strict condition. 
We proposed a modified Brent's method addressing the difficulty of determining the exact value of the noise parameter $\theta$ in the existing studies.
We proved that our proposed mechanism always achieves noise reduction for any given privacy budget $\epsilon>0$, and the noise reduction becomes significant as the privacy budget $\epsilon$ reduces below $1$. 
And we analyzed the variation and optimality of the noise reduction. 
In particular, all the properties still exist in the worst-case $W_1$ mechanism, when the additive noise is largest. 
We showed that the worst-case $W_1$ mechanism is equivalent to the $\ell_1$-sensitivity method, which was first proposed to determine noise for Differential Privacy. We provided both theoretical and empirical analysis in this worst case. 
Experiments on three real-world datasets verify the correctness of all the theoretical results we proposed. This work provides insights for future research aiming to design privacy mechanisms that achieve lower noise and better utility under pufferfish privacy guarantees.


\newpage

\bibliographystyle{ACM-Reference-Format}
\bibliography{sigmod2026}


\newpage

\end{document}